\documentclass[11pt,a4paper]{article}
\usepackage{graphicx} 
\usepackage{amsmath,amsfonts,amssymb,amsthm}
\usepackage{algorithm,algpseudocode}
\usepackage{soul,xcolor}
\usepackage{natbib}
\usepackage{hyperref}
\usepackage{enumitem}
\usepackage{booktabs}

\usepackage[T1]{fontenc}
\usepackage{mathptmx}
\usepackage{microtype}
\usepackage[
  left=2.6cm,
  right=2.6cm,
  top=2.6cm,
  bottom=2.8cm
]{geometry}

\newtheorem{theorem}{Theorem}

\newtheorem{assumption}{Assumption}
\newtheorem{remark}{Remark}

\begin{document}

\title{Sequential Conditional Paired-Sample Testing with Applications to Online Monitoring}
\author{
Xiaoyi Zheng\thanks{Email: \texttt{zxy2000821@mail.ustc.edu.cn}.} \, and  Hong Zhang\thanks{Email: \texttt{zhangh@ustc.edu.cn}.}\\[0.6em]
{\small Department of Statistics and Finance, School of Management, University of Science and Technology}\\
{\small of China, Anhui, China}
}
\date{}
\maketitle

\begin{abstract}
In this paper, we study sequential conditional paired-sample testing, which assesses whether two conditional distributions are equal given a common set of covariates. This problem arises naturally in model monitoring, where the goal is to continuously evaluate discrepancies between predictions and observed labels conditional on covariates. It also arises in other online settings where responses are compared given shared covariates. To address the problem, we propose SCPT, a nonparametric, anytime-valid procedure for sequential conditional paired-sample testing. SCPT uses sequential estimates of a conditional kernel discrepancy to construct a test martingale for continuous monitoring. The procedure provides time-uniform Type I error control and is consistent for a broad class of data-generating processes. We further extend SCPT to accommodate more general dependence structures, comparisons involving multiple auxiliary streams, and conditional goodness-of-fit testing. Simulations and real-data experiments show that SCPT effectively detects differences in conditional distributions in streaming data, demonstrating its practical utility.
\end{abstract}

\section{Introduction}

A fundamental problem in modern statistics is to compare distributions, a task with applications in a wide range of fields. The classical two-sample problem asks whether two independent samples come from the same underlying distribution \citep{hodges1958significance,gretton2012kernel}. In many applications, however, responses depend on observed covariates, and comparing marginal distributions alone may conceal differences at particular covariate values. This motivates conditional distribution testing. One common formulation involves two independent samples with separate covariate observations and potentially different covariate distributions \citep{yan2022distance,zheng2025conditional}. Another concerns paired observations, in which the two responses share the same covariate value and may be dependent \citep{chen2022paired,chatterjee2024kernel}. Both formulations have important practical applications \citep{chang2015nonparametric,chatterjee2024kernel,jin2026testing}.

To date, considerable attention has been devoted to these problems in fixed-sample settings under i.i.d.\ sampling. In applications such as online model monitoring \citep{feng2022clinical}, however, observations arrive sequentially, and the goal is to continuously assess whether the conditional distributions of interest remain equal as data accumulate. It is also often desirable to stop collecting data once sufficient evidence of a discrepancy has emerged, allowing the sample size to depend on the observations. The goal of this paper is therefore to develop a procedure for sequential conditional distribution testing. We focus primarily on the paired setting, in which the two responses share the same covariate value. To formalize the problem, suppose that we observe a stream of triples $\{(X_t,Y_t,Z_t)\}_{t\geq 1}$, where $X_t$ and $Y_t$ are paired responses and $Z_t$ is their shared covariate. Let $\mathcal{F}_{t-1}$ denote the information available from past observations, and let $\mathcal{Z}_0$ be a measurable region of the covariate space. We consider the null hypothesis
\begin{equation}\label{H0:orig}
H_0:\quad
\mathcal{L}(X_t \mid Z_t, \mathcal{F}_{t-1})
=
\mathcal{L}(Y_t \mid Z_t, \mathcal{F}_{t-1})
\quad \text{a.s. on } \{Z_t \in \mathcal{Z}_0\},
\quad \text{for all } t \geq 1.
\end{equation}
Within the region of interest $\mathcal{Z}_0$, $H_0$ requires
$X_t$ and $Y_t$ to have the same distribution given the current
covariate $Z_t$ and the history up to time $t-1$.

The following applications illustrate the scope of \eqref{H0:orig}. Related motivating examples in the fixed-sample i.i.d.\ setting are discussed by \citet{chatterjee2024kernel}; here, we emphasize their use in continuous monitoring. For simplicity, in these examples we take $\mathcal{Z}_0$ to be the full covariate space $\mathcal{Z}$ and assume independence across $t$, so that conditioning on $\mathcal{F}_{t-1}$ in \eqref{H0:orig} can be omitted.

\medskip

\noindent \textit{Sequential Conditional Goodness-of-Fit Testing.}\quad
Goodness-of-fit testing is a classical and extensively studied topic in statistics. Conditional goodness-of-fit (CGOF) testing has received comparatively less attention, despite its importance in many applications. Given a response $Y_t\in\mathbb{R}^q$ and covariates $Z_t$, let $F(\cdot\mid z)$ be a specified conditional cumulative distribution function induced by a target model. Sequential CGOF testing aims to assess whether, at every time $t\geq 1$, the conditional distribution of $Y_t$ given $Z_t$ coincides with the target distribution:
\[
Y_t\mid Z_t=z \sim F(\cdot\mid z).
\]
Existing approaches for CGOF testing include tests based on kernel
smoothing \citep{zheng2012testing}, kernel Stein discrepancy
\citep{liu2016kernelized,jitkrittum2020testing,baum2023kernel}, and several other techniques.
Most existing methods are designed for fixed-sample inference under
i.i.d.\ sampling, whereas applications such as model monitoring
motivate sequential formulations
\citep{feng2022clinical,cho2026gaavi}. Sequential CGOF testing fits into our framework by taking the conditional distribution of $X_t$ given $Z_t$ to be known and fixed over time:
\[
X_t\mid Z_t=z\sim F(\cdot\mid z).
\]
Specifically, at each time $t$, one can generate an auxiliary observation $X_t$ from $F(\cdot\mid Z_t)$, independently of $Y_t$ conditional on $Z_t$. The CGOF hypothesis then becomes the equality of the conditional distributions of $X_t$ and $Y_t$ given $Z_t$, as formulated in \eqref{H0:orig}. 

\medskip

\noindent \textit{Monitoring Model Calibration.}\quad
Monitoring the calibration of predictive models is a natural application of sequential CGOF testing. Well-calibrated predictions support reliable decision-making in many applications \citep{Guo2017calibration}, making it useful to assess whether calibration is maintained as new observations arrive \citep{feng2022clinical}. For example, in multiclass classification, a predictive model outputs a probability vector $p(z)=\{p_y(z)\}_{y\in\mathcal{Y}}$, where $\mathcal{Y}$ is the set of class labels and $p_y(z)$ denotes the predicted probability of class $y$ given the covariate value $z$. Let $Y_t\in\mathcal{Y}$ and $Z_t$ denote the true class label and the covariates observed at time $t$, respectively. Multiclass calibration requires that, for any $t\geq 1$,
\[
\mathbb{P}\left\{Y_t=y\mid p(Z_t)\right\}=p_y(Z_t), \quad \forall\, y\in\mathcal{Y}.
\]
This condition defines a sequential CGOF problem. Similar calibration tests can be formulated for other predictive models, including quantile regression and generative models. The i.i.d.\ formulations discussed by \citet{chatterjee2024kernel} extend naturally to the sequential framework considered here.

\medskip

\noindent \textit{Monitoring Conditional Generative Models.}\quad
Another application is monitoring whether a conditional generative model produces outputs whose conditional distributions coincide with those of the observed responses. When the model-induced conditional distribution is available in a closed form, this is a sequential CGOF problem. For many modern generative models, however, including variational autoencoders (VAEs), generative adversarial networks (GANs), diffusion models, and large language models (LLMs), a tractable or fully accessible representation of this distribution may not be available.

Model-generated samples nevertheless allow us to assess such models within the same framework. We consider a setting in which a single model-generated output $X_t$ and a corresponding ground-truth response $Y_t$ are available for each observed covariate value $Z_t$. Testing whether $\mathbb{P}_{X_t\mid Z_t}$ and $\mathbb{P}_{Y_t\mid Z_t}$ coincide then gives precisely the problem in \eqref{H0:orig}. When multiple outputs can be generated for each covariate value, the additional samples may improve power. We discuss this extension, together with the use of a known reference distribution in CGOF testing, in Section~\ref{cgof_test}.

\medskip

\noindent \textit{Online Paired Comparisons.}\quad
Our framework also applies to online comparisons of paired responses from two systems or populations sharing the same covariates. A typical example is {shadow deployment} in production applications \citep{schermann2018we,satyal2018shadow}. Before replacing an existing model or software system, a candidate version is run alongside the current production version on the same inputs. The candidate's outputs are not exposed to end users, allowing the two versions to be compared without changing the user-facing service. Let $X_t$ and $Y_t$ denote the responses of the current and candidate versions, respectively, to input $Z_t$. The goal is to continuously assess whether their conditional response distributions coincide, as formulated in \eqref{H0:orig}.

Similar comparisons arise when evaluating systems operating under common environmental conditions. For example, \citet{schoeneberger2026multisensor} collected temperature and humidity measurements using co-located low-cost and professional sensors. A practical motivation for comparing these instruments is to assess whether affordable sensors can provide measurements consistent with professional references, supporting their use in distributed environmental monitoring networks. Here, $X_t$ and $Y_t$ denote the two sensors' readings of temperature or relative humidity, while $Z_t$ contains shared environmental covariates, such as incoming shortwave and longwave radiation. Comparing their conditional distributions can reveal systematic discrepancies even when the readings are highly correlated overall. Since sensor measurements are collected over time, sequential
testing provides a natural way to detect such discrepancies
as evidence accumulates.
\medskip

The motivating examples above call for the development of sequential methods for conditional testing, and the statistical framework of \emph{anytime-valid inference} provides tools for this purpose \citep{ramdas2023game}.
Within this framework, \emph{test-by-betting} methods construct evidence processes through bets against the null hypothesis \citep{peter2019safe,waudby2024estimating}. In particular, \citet{shekhar2023nonparametric} developed nonparametric sequential two-sample tests, and \citet{podkopaev2023sequential} proposed sequential tests of independence. Test-by-betting methods have also been used for fairness auditing \citep{chugg2023auditing} and stochastic dominance testing \citep{arnold2026betting}, among other applications. Nevertheless, much of this work concerns global, unconditional hypotheses, with theoretical analyses typically assuming independent and often identically distributed observations.

In this paper, we develop SCPT, a nonparametric procedure for sequential conditional paired-sample testing. SCPT combines conditional kernel maximum mean discrepancy, nearest-neighbor estimation, and test martingales to test the null hypothesis in \eqref{H0:orig}. It provides finite-sample, anytime-valid Type I error control and is consistent against a broad class of alternatives under suitable conditions, accommodating certain forms of temporal dependence and nonstationarity. We further develop extensions for more general dependence structures, mixtures of hyperparameter configurations, one-versus-many comparisons, and CGOF testing. Under suitable conditions, the approach also extends to settings in which
the two samples have separate covariate observations, as discussed in
Section~\ref{conclusion}. Extensive simulations and two real-data examples demonstrate the practical utility of SCPT for online monitoring. The resulting framework connects conditional two-sample testing with anytime-valid inference and supports applications such as online monitoring of conditional generative models and conditional density models, a setting in which sequential testing has received limited attention.

\section{Preliminaries}\label{background}
We first introduce some notation. Suppose we observe a stream $\{(X_t,Y_t,Z_t)\}_{t\geq 1}$, where $X_t,Y_t\in\mathcal{X}$ and $Z_t\in\mathcal{Z}$. For simplicity, we assume that $\mathcal{Z}=\mathbb{R}^p$ and that $\mathcal{X}$ is a closed subset of $\mathbb{R}^q$, with $p,q\geq 1$. Categorical responses can be accommodated by encoding their values as positive integers. More generally, the testing problem and the proposed methods can be formulated on Polish spaces, although we do not pursue this extension in the present paper. Let $\mathcal{F}_t=\sigma\bigl((X_s,Y_s,Z_s):1\leq s\leq t\bigr)$ denote the natural filtration, with $\mathcal{F}_0$ the trivial $\sigma$-field. We let $\mathcal{Z}_0\subseteq\mathcal{Z}$ be a measurable region of interest and write $\mathcal{S}_t$ for the support of $Z_t$.

In the Euclidean setting, we aim to sequentially test the null hypothesis
\begin{equation}\label{H0:1}
\mathbb{P}(X_t\leq u\mid Z_t,\mathcal{F}_{t-1})
=
\mathbb{P}(Y_t\leq u\mid Z_t,\mathcal{F}_{t-1})
\quad \text{on }\{Z_t\in\mathcal Z_0\}, \quad
\text{for any} ~ t\geq 1, u\in\mathbb{R}^q.
\end{equation}
Here and throughout the paper, unless otherwise stated, all equalities, inequalities, and convergence statements involving random variables are understood to hold almost surely; we omit this qualifier for brevity.

We write $\mathbb{P}_{Z_t}$ for the marginal distribution of $Z_t$ and $\mathbb{P}_{X_t\mid Z_t}$, $\mathbb{P}_{Y_t\mid Z_t}$, and $\mathbb{P}_{(X_t,Y_t)\mid Z_t}$ for the regular conditional distributions of $X_t$, $Y_t$, and $(X_t,Y_t)$ given $Z_t$, respectively. Analogous notation is used for other marginal and conditional distributions.

\medskip

\noindent \textit{RKHS and Kernel Maximum Mean Discrepancy.}\quad  
A {reproducing kernel Hilbert space} (RKHS) $\mathcal{H}$ is a Hilbert space of functions on a domain $\mathcal{X}$ associated with a positive definite kernel $\mathrm{K}:\mathcal{X}\times \mathcal{X}\to \mathbb{R}$ and endowed with an inner product $\langle\cdot, \cdot\rangle_\mathcal{H}$. The key property of $\mathcal{H}$ is the {reproducing property}:
\[
f(x) = \langle f, \mathrm{K}(x, \cdot) \rangle_{\mathcal{H}}, \quad \forall f \in \mathcal{H}, \, x \in \mathcal{X},
\]
which allows the evaluation of a function at any point via the inner product in the Hilbert space. RKHS embeddings of probability distributions exploit this property by mapping a random variable $X \sim \mathbb{P}_X$ to its {kernel mean embedding} 
\[
\mu_X = \mathbb{E}\left[\mathrm{K}(X, \cdot)\right] \in \mathcal{H},
\] 
which can further be viewed as a mapping from $\mathcal{P}(\mathcal{X})$, the set of Borel probability measures on $\mathcal{X}$, to $\mathcal{H}$. For characteristic kernels, such as the Gaussian and Laplacian kernels, the corresponding kernel mean embedding uniquely characterizes the underlying distribution. This property enables nonparametric measures of distributional discrepancy, most notably the {maximum mean discrepancy} (MMD):
\[
\mathrm{MMD}^2(\mathbb{P}_X, \mathbb{P}_Y) 
= \big\|\mu_X - \mu_Y\big\|_{\mathcal{H}}^2
= \mathbb{E}\left[\mathrm{K}(X,X')\right] + \mathbb{E}\left[\mathrm{K}(Y,Y')\right] - 2 \mathbb{E}\left[\mathrm{K}(X,Y)\right],
\]
where $X,X'\sim \mathbb{P}_X$ and $Y,Y'\sim \mathbb{P}_Y$ are independent. The MMD is non-negative and equals zero if and only if $\mathbb{P}_X = \mathbb{P}_Y$ \citep{gretton2012kernel}.
\medskip

\noindent \textit{Test Martingales and Anytime-Valid Inference.}\quad A powerful tool for sequential testing of distributional hypotheses is the {test martingale} \citep{ramdas2023game,peter2019safe,waudby2024estimating}. A sequence of random variables $\{M_t\}_{t\ge 0}$ is called a test martingale for the null hypothesis $H_0$ if $M_0\leq1$, $M_t \ge 0$, and $\mathbb{E}[M_{t+1}\mid \mathcal{F}_t] = M_t$ under $H_0$, where $\mathcal{F}_t$ denotes the filtration generated by the observations up to time $t$. The test martingale $M_t$ can be interpreted as the wealth of a gambler playing a game at time $t$. Under $H_0$, the game is ``fair,'' making it very unlikely for the gambler to eventually become a millionaire. This intuition is formally captured by Ville's inequality \citep{ville1939}: for any positive martingale $M_t$ with $\mathbb{E}[M_0] \leq 1$ and any constant $0<\alpha<1$, we have
\begin{equation}\label{ville}
\mathbb{P}\left(\exists \, t \geq 1 \text{ such that } M_t > \alpha^{-1}\right)\leq \alpha.
\end{equation}
Therefore, rejecting the null hypothesis as soon as $M_t>1/\alpha$
yields an anytime-valid, level-$\alpha$ sequential test: we may inspect $M_t$ at every time $t$ and stop when it crosses
the threshold, while the probability of ever falsely rejecting
a null hypothesis remains at most $\alpha$.

A standard construction of a test martingale is the {capital process} \citep{shekhar2023nonparametric,waudby2024estimating}. Suppose that we
wish to test a null hypothesis \(H_0\) sequentially based on i.i.d. observations
\(\{X_i\}_{i=1}^{\infty}\). Let
\(
\mathcal F_{t-1}
=
\sigma(X_1,\ldots,X_{t-1})
\)
denote the information available before time \(t\). At each time \(t\),
one selects an \(\mathcal F_{t-1}\)-measurable function
\(f_t(\cdot)\) and betting fraction \(\lambda_t\) such that
\(
\lambda_t f_t(X_t)>-1
\) and
\(
\mathbb E\left[
f_t(X_t)\mid\mathcal F_{t-1}
\right]
=
0
\) under \(H_0\).
We call \(S_t=f_t(X_t)\) the ``payoff'' at time \(t\).
Starting with \(M_0=1\), we define the capital process by
\begin{equation}\label{capital}
M_t
=
\prod_{s=1}^{t}
\left\{
1+\lambda_s S_s
\right\},
\end{equation}
which can be updated sequentially using the recursion \(M_{t+1}=M_t\{1+\lambda_{t+1}S_{t+1}\}\). Intuitively, \(M_t\) represents the wealth of a gambler who sequentially
bets on each new observation. Under \(H_0\),
\[
\mathbb E[M_t\mid\mathcal F_{t-1}]
=
M_{t-1}
\left\{
1+\lambda_t
\mathbb E[f_t(X_t)\mid\mathcal F_{t-1}]
\right\}
=
M_{t-1},
\]
so \(\{M_t\}_{t\geq 0}\) is a test martingale.

\section{Method}\label{method}
In this section, we present the proposed method for sequential conditional paired-sample testing, which we refer to as {SCPT} throughout the paper. We begin by introducing several mathematical assumptions, then derive the proposed procedure and establish its theoretical properties. Finally, we discuss several useful extensions.

\subsection{Construction of the Test}\label{construction}
Our goal is to construct a test martingale for the hypothesis testing problem in \eqref{H0:1}. The standard construction can be summarized as follows: at each time $t$, we first use conditional kernel mean embeddings to define a ``payoff'' $S_t$ satisfying $-1 \leq S_t \leq 1$ and $\mathbb{E}[S_t \mid Z_t, \mathcal{F}_{t-1}] = 0$ under the null (Sections~\ref{assumption}--\ref{knn}). We then form a test martingale analogous to the capital process in \eqref{capital} by carefully choosing the betting fraction $\lambda_t$ (Section~\ref{lbow}).

\subsubsection{Assumptions on Dependence and Kernels}\label{assumption}
Although the proposed framework can accommodate more general dependence structures, the filtration $\mathcal{F}_{t-1}$ appearing in the null hypothesis makes it complicated to define the test and to analyze its theoretical properties under the alternative. Therefore, throughout the construction of the test martingale, we impose the following conditional independence assumption (Assumption~\ref{assumption_independence}) for clarity of exposition. We defer the extension to such general dependence structures to Section~\ref{general_dependence}.

\begin{assumption}\label{assumption_independence} For each $t\geq 1$, the joint distribution $\mathbb{P}_{1:t}$ of
$\{(X_s,Y_s,Z_s)\}_{s=1}^{t}$ admits the decomposition
\[
d\mathbb{P}_{1:t}
=
d\mathbb{P}_{(Z_1,\ldots,Z_t)}
\prod_{s=1}^{t}
d\mathbb{P}_{(X_s,Y_s)\mid Z_s},
\]
where $\mathbb{P}_{(Z_1,\ldots,Z_t)}$ denotes the joint distribution of
$(Z_1,\ldots,Z_t)$.
\end{assumption}

Equivalently, conditional on $(Z_1,\ldots,Z_t)$, the pairs
$(X_s,Y_s)$, $s=1,\ldots,t$, are mutually independent, and the conditional
distribution of  $(X_s,Y_s)$ depends on the covariate sequence only through
the current covariate $Z_s$. This assumption is
weaker than requiring the full sequence
$\{(X_s,Y_s,Z_s)\}_{s=1}^{\infty}$ to be independent. Under Assumption~\ref{assumption_independence}, the original null hypothesis is equivalent to
\begin{equation}\label{H0:iid}
H_0:\, 
\mathbb{P}(X_t\leq u \mid Z_t=z)
=
\mathbb{P}(Y_t\leq u \mid Z_t=z),
\,
\forall\, u\in\mathbb{R}^q,\ 
z\in\mathcal{Z}_0\cap\mathcal{S}_t,\ 
t\geq 1,
\end{equation}
where the filtration $\mathcal{F}_{t-1}$ no longer appears. We now proceed to construct the test martingale. Let $\mathcal{H}$ be a separable RKHS on $\mathcal{X}$, and let $\mathrm{K}(\cdot,\cdot):\mathcal{X}\times\mathcal{X}\to\mathbb{R}$ denote its associated kernel function. To ensure the validity of the proposed approach, we impose the following assumption on $\mathrm{K}$.

\begin{assumption}\label{assumption_kernel}
 The kernel $\mathrm{K}: \mathcal{X}\times\mathcal{X} \rightarrow \mathbb{R}$ satisfies the following conditions:
\begin{enumerate}[label=(\roman*)]
\item $\mathrm{K}$ is measurable and positive definite on $\mathcal{X}\times\mathcal{X}$, and $0\leq \mathrm{K}(x, y)\leq 1$ for all $(x,y)\in\mathcal{X}\times\mathcal{X}$.
\item The kernel $\mathrm{K}$ is characteristic, that is, the mean embedding $\mu: \mathcal{P}(\mathcal{X}) \rightarrow \mathcal{H}$ defined by $\mu(\mathbb{P})=\mathbb{E}_{X\sim\mathbb{P}}\left[\mathrm{K}(X,\cdot)\right]$ is injective, where $\mathcal{P}(\mathcal{X}) $  denotes the collection of all Borel probability measures on $\mathcal{X}$.
\end{enumerate}
\end{assumption}

We next present several examples of kernel functions satisfying Assumption~\ref{assumption_kernel}, whose characteristic property follows from the results of \citet{sriperumbudur2010hilbert}.

\medskip
\noindent\textit{Dirac Kernel.}\quad For categorical $X$ and $Y$ taking values in the same set $\mathcal{X}$, the Dirac kernel is defined as
$$
\mathrm{K}(x,y)=\mathbb{I}(x=y),
$$
where $\mathbb{I}(\cdot)$ denotes the indicator function.

\medskip
\noindent\textit{Gaussian Kernel.}\quad The Gaussian kernel is defined as
$$
\mathrm{K}(x,y)=\exp\left\{-\frac{\|x-y\|_2^2}{2\sigma^2}\right\},
$$
where $\sigma>0$ is a bandwidth parameter.

\medskip
\noindent\textit{Laplacian Kernel.}\quad The Laplacian kernel is defined as
$$
\mathrm{K}(x,y)=\exp\left\{-\frac{\|x-y\|_2}{\sigma}\right\},
$$
where $\sigma>0$ is a bandwidth parameter.

\subsubsection{Reduction to Mean Testing via RKHS Embedding}\label{define_payoff}
For each observation $(X_t,Y_t,Z_t)$, to quantify the difference between the conditional distributions, we define the conditional kernel mean embeddings $\mu_{X_t\mid Z_t}$ and $\mu_{Y_t\mid Z_t}$ by
$$
\mu_{X_t\mid Z_t}=\mathbb{E}\left[\mathrm{K}(X_t,\cdot)\mid Z_t\right]\text{~and~} \mu_{Y_t\mid Z_t}=\mathbb{E}\left[\mathrm{K}(Y_t,\cdot)\mid Z_t\right],
$$
respectively, which are $\mathcal{H}$-valued random variables. We also use the notation
$$
\mu_{X_t\mid Z_t=z}=\mathbb{E}\left[\mathrm{K}(X_t,\cdot)\mid Z_t=z\right]\text{~and~} \mu_{Y_t\mid Z_t=z}=\mathbb{E}\left[\mathrm{K}(Y_t,\cdot)\mid Z_t=z\right]
$$ 
for each $z\in \mathcal{S}_t$. This approach has been employed in several previous works in non-sequential settings \citep{fukumizu2007kernel,ren2016conditional,yan2022distance,chatterjee2024kernel,moskvichev2026measuring}. We then define the following squared conditional MMD at $Z_t=z$ with $z\in \mathcal{S}_t$ as
$$
\begin{aligned}\label{LMMD}
    \delta_t(z):=~&\|\mu_{X_t\mid Z_t=z}-\,\mu_{Y_t\mid Z_t=z}\|_{\mathcal{H}}^2\\
    =~& \mathbb{E}_{(X,Y),(X',Y')\sim\mathbb{P}_{(X_t,Y_t)\mid Z_t=z}}\left[\mathrm{K}(X,X')+\mathrm{K}(Y,Y')-\mathrm{K}(X,Y')-\mathrm{K}(Y,X')\right],
\end{aligned}
$$
where $(X,Y)$ and $(X',Y')$ are two independent copies drawn from
$\mathbb{P}_{(X_t,Y_t)\mid Z_t=z}$. By Assumption~\ref{assumption_kernel}, $\mathbb{P}_{X_t\mid Z_t}=\mathbb{P}_{Y_t\mid Z_t}$ on $\mathcal{Z}_0$ is equivalent to 
\(\delta_t(z)=0\) for any \(z\in\mathcal{Z}_0\cap\mathcal{S}_t\). Motivated by this observation, for a given 
weight function \(w:\mathbb{R}^p\to\mathbb{R}_{+}\), we define the aggregated 
conditional MMD $\bar{\delta}_t(w)$ as
\begin{equation}\label{ELMMD}
\bar{\delta}_t(w)=\mathbb{E}\left[\delta_t(Z_t)w(Z_t)\right].
\end{equation}
We impose the following assumption on the weight function, under which the 
equivalence between $\mathbb{P}_{X_t\mid Z_t}=\mathbb{P}_{Y_t\mid Z_t}$ on $\mathcal{Z}_0$ and 
\(\bar{\delta}_t(w)=0\) follows immediately. 

\begin{assumption}\label{weight}
The weight function \(w(\cdot)\) is measurable and satisfies \(0\leq w(\cdot)\leq 1\), \(w(z)=0\) for all \(z\in\mathcal{Z}_0^c\), and
there exists a constant \(w_0>0\) such that \(w(z)\geq w_0\) for all
\(z\in\mathcal{Z}_0\). 
\end{assumption}

A natural choice satisfying this assumption is the indicator weight
\(w(z)=\mathbb{I}\{z\in\mathcal{Z}_0\}\). In particular, when
\(w\equiv1\) and the observations are i.i.d., \(\bar{\delta}_t(w)\)
coincides with the squared expected conditional maximum mean discrepancy
(ECMMD) introduced by \citet{chatterjee2024kernel}, who studied its
estimation in the fixed-sample setting.
In our sequential setting, we instead construct the test using the following zero-mean characterization: for any \(t\geq 1\),
\begin{equation}\label{MMD}
\mathbb{E}\left[
\left\langle
a(Z_t;\cdot),
w(Z_t)\{\mathrm{K}(X_t,\cdot)-\mathrm{K}(Y_t,\cdot)\}
\right\rangle_{\mathcal{H}}
\right]=0
\end{equation}
for any measurable function \(z\mapsto a(z;\cdot)\in\mathcal{H}\) satisfying \(\|a(z;\cdot)\|_{\mathcal{H}}\leq 1\) for all \(z\in\mathcal{S}_t\). This condition is equivalent to the null hypothesis \eqref{H0:iid}; a rigorous proof is provided in Section~S1.1 of the Supplementary Material. Under the alternative hypothesis, we seek a function that maximizes the left-hand side of \eqref{MMD}, referred to as the ``witness function'', and construct the test based on this function. This strategy is similar in spirit to existing test-martingale-based sequential tests \citep{shekhar2023nonparametric,podkopaev2023sequential}, which begin with a measure of departure from the null that admits a variational representation, use the representation to derive a witness function, and then construct a capital process as in \eqref{capital}.

By the Cauchy--Schwarz inequality, a witness function for \eqref{MMD}  is
\begin{equation}\label{witness}
a^*(Z_t;\cdot)
=
\frac{
\mathbb{E}[\mathrm{K}(X_t,\cdot)\mid Z_t]
-
\mathbb{E}[\mathrm{K}(Y_t,\cdot)\mid Z_t]
}{
\left\|
\mathbb{E}[\mathrm{K}(X_t,\cdot)\mid Z_t]
-
\mathbb{E}[\mathrm{K}(Y_t,\cdot)\mid Z_t]
\right\|_{\mathcal{H}}
},
\end{equation}
where we set $a^*(Z_t;\cdot)=0$ whenever the denominator vanishes.
Motivated by the form of the witness function, we construct the following test process, analogous to the capital process in \eqref{capital}:
\begin{equation}\label{martingale}
M_t
=
\prod_{s=1}^{t}
\left\{
1+
\lambda_s(Z_s)\frac{w(Z_s)}{\sqrt{2}}
\frac{
\left\langle
U_s(Z_s,\cdot)-V_s(Z_s,\cdot),\,
\mathrm{K}(X_s,\cdot)-\mathrm{K}(Y_s,\cdot)
\right\rangle_{\mathcal{H}}
}{
\max\{\left\|U_s(Z_s,\cdot)-V_s(Z_s,\cdot)\right\|_{\mathcal{H}},\eta\}
}
\right\}.
\end{equation}
In \eqref{martingale}, $\eta>0$ is a small truncation parameter introduced to avoid degeneracy of the denominator; \(U_t(Z_t,\cdot)\) and \(V_t(Z_t,\cdot)\) denote estimators of the conditional mean embeddings
\[
\mathbb{E}\!\left[\mathrm{K}(X_t,\cdot)\mid Z_t\right]
\quad\text{and}\quad
\mathbb{E}\!\left[\mathrm{K}(Y_t,\cdot)\mid Z_t\right],
\]
respectively, constructed using only observations available before time \(t\). Consequently, \(U_t(\cdot,\cdot)\) and \(V_t(\cdot,\cdot)\) are \(\mathcal{F}_{t-1}\)-measurable random functions. 
The betting fraction \(\lambda_t(\cdot)\) takes values in \((-1,1)\) and is \(\mathcal{F}_{t-1}\)-measurable for each \(t\geq1\). Because the test concerns conditional distributions, the betting fraction at time \(t\) is allowed to depend on the current covariate \(Z_t\). Accordingly, \(\lambda_t(\cdot)\) is a covariate-indexed function rather than the scalar  used in the standard capital process \eqref{capital}. The normalizing factor \(\sqrt{2}\) ensures that the scaled payoff in each multiplicative factor lies in \([-1,1]\). Indeed, the Cauchy--Schwarz inequality gives
\[
\begin{aligned}
&\left|
\frac{
\left\langle
U_t(Z_t,\cdot)-V_t(Z_t,\cdot),\,
\mathrm{K}(X_t,\cdot)-\mathrm{K}(Y_t,\cdot)
\right\rangle_{\mathcal{H}}
}{
\max\{\left\|U_t(Z_t,\cdot)-V_t(Z_t,\cdot)\right\|_{\mathcal{H}},\eta\}
}
\right| \\
\leq&
\left\|\mathrm{K}(X_t,\cdot)-\mathrm{K}(Y_t,\cdot)\right\|_{\mathcal{H}} \\
=&
\sqrt{
\mathrm{K}(X_t,X_t)+\mathrm{K}(Y_t,Y_t)
-\mathrm{K}(X_t,Y_t)-\mathrm{K}(Y_t,X_t)
} \\
\leq & \sqrt{2}.
\end{aligned}
\]
Consequently, because \(\lambda_t(Z_t)\in(-1,1)\), every multiplicative factor in \eqref{martingale} is strictly positive. 

In what follows, we provide explicit constructions of \(U_t(Z_t,\cdot)\), \(V_t(Z_t,\cdot)\), and \(\lambda_t(Z_t)\) for implementing the test process, beginning with the conditional mean embedding estimators \(U_t(Z_t,\cdot)\) and \(V_t(Z_t,\cdot)\) and then turning to the betting fraction \(\lambda_t(Z_t)\).

\subsubsection{Choice of Conditional Mean Estimators}\label{knn}

  We employ a sequential k-nearest neighbor (KNN) approach to estimate the conditional expectations $\mathbb{E}[\mathrm{K}(X_t,\cdot)\mid Z_t]$ and $\mathbb{E}[\mathrm{K}(Y_t,\cdot)\mid Z_t]$.  Let $k=k(t)$ with $k(t)=o(t)$, $k(t)\rightarrow\infty$, and let $\mathcal{N}_t(k(t))$ be the index set of the $k(t)$-nearest neighbors of $Z_t$ among $(Z_1,\ldots,Z_{t-1})$ calculated using standard Euclidean distance. We then set:
$$
U_{t}(Z_t,\cdot)=\frac{1}{k(t)}\sum_{i\in \mathcal N_{t}(k(t))} \mathrm{K}(X_i,\cdot),\quad
V_{t}(Z_t,\cdot)=\frac{1}{k(t)}\sum_{i\in \mathcal N_{t}(k(t))} \mathrm{K}(Y_i,\cdot).
$$
By the reproducing property of the RKHS, a direct calculation yields
\begin{align}\label{calc}
    &\left\langle U_t(Z_t,\cdot)-V_t(Z_t,\cdot),\,\mathrm{K}(X_t,\cdot)-\mathrm{K}(Y_t,\cdot)\right\rangle_{\mathcal{H}}\notag\\
    = & \frac{1}{k(t)}\sum_{i\in \mathcal N_{t}(k(t))}
    \left[
    \mathrm{K}(X_i,X_t)+\mathrm{K}(Y_i,Y_t)-\mathrm{K}(X_i,Y_t)-\mathrm{K}(Y_i,X_t)
    \right],
\end{align}
\begin{equation}\label{calc2}
    \| U_t(Z_t,\cdot)-V_t(Z_t,\cdot)\|_{\mathcal{H}}
    = \frac{1}{k(t)}\left(\sum_{i,j\in \mathcal N_{t}(k(t))}
    \left[
    \mathrm{K}(X_i,X_j)+\mathrm{K}(Y_i,Y_j)-\mathrm{K}(X_i,Y_j)-\mathrm{K}(Y_i,X_j)
    \right]\right)^{1/2}.
\end{equation}
 To simplify notation, define
\begin{equation}\label{eq:tilde-S}
\widetilde{S}_t
=
\frac{
\sum_{i\in\mathcal N_t(k(t))}
\left[
\mathrm{K}(X_i,X_t)
+\mathrm{K}(Y_i,Y_t)
-\mathrm{K}(X_i,Y_t)
-\mathrm{K}(Y_i,X_t)
\right]
}{
\sqrt{2}
\max\left\{
\left(
\sum_{i,j\in\mathcal N_t(k(t))}
\left[
\mathrm{K}(X_i,X_j)
+\mathrm{K}(Y_i,Y_j)
-\mathrm{K}(X_i,Y_j)
-\mathrm{K}(Y_i,X_j)
\right]
\right)^{1/2},
\eta k(t)
\right\}
}.
\end{equation}
Let
\(
S_t=w(Z_t)\widetilde{S}_t.
\)
By \eqref{calc} and \eqref{calc2}, the test process can be written as $M_t
=
\prod_{s=1}^{t}
\left\{
1+\lambda_s(Z_s)S_s
\right\}$, where \(S_t\) plays the role of the payoff in the standard capital
process \eqref{capital}.

\subsubsection{Choice of the Betting Fractions}\label{lbow}

A standard strategy for selecting the betting function
$\lambda_t(\cdot)$, and hence the betting fraction $\lambda_t(Z_t)$,
is to maximize the expected one-step log growth of the test martingale.
Specifically, at each time $t$, one seeks to solve
\begin{equation}\label{argmax}
\lambda_t(\cdot)
\in
\operatorname*{arg\,max}_{\lambda\in\Lambda}
\mathbb E\left[
\log\{1+\lambda(Z_t)S_t\}
\right],
\end{equation}
where $\Lambda$ denotes a candidate class of betting functions.
Because direct optimization over $\Lambda$ is difficult, we use a
lower bound on the objective functional to derive a simple,
easily implemented betting rule. Let
\[
\Lambda
=
\left\{
\lambda:\mathcal Z\to[0,1):
\lambda\text{ is measurable}
\right\}
\]
denote the class of candidate betting functions. The nonnegativity
constraint is motivated by the direction of the oracle witness
function, whose conditional expected payoff is non-negative. Consider the population expected
log-growth functional
\[
\mathcal L_t(\lambda)
=
\mathbb E\left[
\log\{1+\lambda(Z_t)S_t\}
\right],
\qquad
\lambda\in\Lambda.
\]
Using the inequality \(\log(1+\lambda x)\geq \lambda x+x^2\{\log(1-\lambda)+\lambda\}\), valid for all $\lambda\in[0,1)$ and $x\in[-1,1]$, we obtain
\begin{equation}\label{quadratic_approximation}
\begin{aligned}
\mathcal L_t(\lambda)
&\geq
\mathbb E\left[
\lambda(Z_t)S_t+S_t^2\left\{\log(1-\lambda(Z_t))+\lambda(Z_t)\right\}
\right]\\
&=
\mathbb E\left[
\lambda(Z_t)\mathbb E[S_t\mid Z_t]
+\left\{\log(1-\lambda(Z_t))+\lambda(Z_t)\right\}
\mathbb E[S_t^2\mid Z_t]
\right].
\end{aligned}
\end{equation}
We maximize this lower bound pointwise in $Z_t$. If $\mathbb E[S_t\mid Z_t]\leq 0$, the conditional expression is non-positive and is maximized at $\lambda(Z_t)=0$. If $\mathbb E[S_t\mid Z_t]>0$, elementary calculus yields the maximizer
\begin{equation}\label{maximizer_lambda}
\lambda(Z_t)=
\frac{\mathbb E[S_t\mid Z_t]}
{\mathbb E[S_t\mid Z_t]+\mathbb E[S_t^2\mid Z_t]}.
\end{equation}
Since \(S_t=w(Z_t)\widetilde S_t\), we have
\[
\mathbb E[S_t\mid Z_t]
=
w(Z_t)\mathbb E[\widetilde S_t\mid Z_t],
\quad
\mathbb E[S_t^2\mid Z_t]
=
w^2(Z_t)\mathbb E[\widetilde S_t^2\mid Z_t].
\]
We then use past observations to estimate the first and second moments of \(\widetilde S_t\) conditional on \(Z_t\).
We compute a leave-one-out payoff for each pair in the current
neighborhood using a witness function fitted to the remaining pairs.
Specifically, for $k(t)\geq 2$ and $i\in\mathcal{N}_t(k(t))$, define
\begin{equation}\label{eq:loo-payoff}
\begin{aligned}
&\widehat d_t^{(-i)}
=
\frac{1}{k(t)-1}
\sum_{j\in\mathcal{N}_t(k(t))\setminus\{i\}}
\bigl[\mathrm K(X_j,\cdot)-\mathrm K(Y_j,\cdot)\bigr],\qquad
\widetilde S_{t,i}^{(-i)}\\
&=
\frac{
\displaystyle
\sum_{j\in\mathcal{N}_t(k(t))\setminus\{i\}}
\bigl[
\mathrm K(X_j,X_i)+\mathrm K(Y_j,Y_i)
-\mathrm K(X_j,Y_i)-\mathrm K(Y_j,X_i)
\bigr]
}{
\sqrt{2}\max\!\left\{
\left(
\sum_{j,\ell\in\mathcal{N}_t(k(t))\setminus\{i\}}
\bigl[
\mathrm K(X_j,X_\ell)+\mathrm K(Y_j,Y_\ell)
-\mathrm K(X_j,Y_\ell)-\mathrm K(Y_j,X_\ell)
\bigr]
\right)^{1/2},
\,(k(t)-1)\eta
\right\}
}.
\end{aligned}
\end{equation}
For $k(t)=1$, set $\widetilde S_{t,i}^{(-i)}=0$.
Substituting the KNN estimates
\[
u_t=
\frac{1}{k(t)}\sum_{i\in\mathcal{N}_t(k(t))}
\widetilde S_{t,i}^{(-i)},
\qquad
v_t=
\frac{1}{k(t)}\sum_{i\in\mathcal{N}_t(k(t))}
\bigl(\widetilde S_{t,i}^{(-i)}\bigr)^2
\]
into \eqref{maximizer_lambda} gives
\begin{equation}\label{eq:betting-function}
\lambda_t=
\begin{cases}
\displaystyle\frac{u_t}{u_t+w(Z_t)v_t},
& u_t>0 \text{ and } w(Z_t)>0,\\[4pt]
0, & \text{otherwise}.
\end{cases}
\end{equation}
Algorithm~\ref{alg:approx} summarizes the resulting SCPT procedure.

\begin{remark}
Since $v_t\geq u_t^2$, \eqref{eq:betting-function} gives $0\leq\lambda_t<1$. Hence every factor $1+\lambda_t S_t$ is strictly positive, as is $M_t$.
\end{remark}

\begin{remark}
Our construction of $\{\lambda_t\}_{t\geq1}$ is similar to the
lower bound on wealth (LBOW) approach used by
\citet{waudby2024estimating} for unconditional mean testing.
Here, we adapt this approach to conditional testing through
covariate-dependent betting functions.
\end{remark}

\begin{algorithm}[t]
\caption{Sequential Conditional Paired-Sample Testing (SCPT)}
\label{alg:approx}
\begin{algorithmic}[1]
\State \textbf{Input:} Kernel function $\mathrm{K}$, KNN neighborhood numbers $k(t)$, weight function $w$, significance level $\alpha$, truncation parameter $\eta$;
\State Observe $(X_1,Y_1,Z_1)$. Initialize $M_0=M_1=1$, $\lambda_1=\widetilde S_1=S_1=0$;
\For{$t = 2,3,\ldots$}
    \State Observe $(X_t,Y_t,Z_t)$.
    \State Obtain neighborhood set $\mathcal{N}_t(k(t))$ for $Z_t$ based on $\{Z_1,\ldots,Z_{t-1}\}$;
    \State For each $i\in\mathcal{N}_t(k(t))$, compute
$\widetilde S_{t,i}^{(-i)}$ using \eqref{eq:loo-payoff} if $k(t)\geq 2$,
and set $\widetilde S_{t,i}^{(-i)}=0$ if $k(t)=1$.
    \State Set $u_t=k(t)^{-1}\sum_{i\in\mathcal{N}_t(k(t))}\widetilde S_{t,i}^{(-i)}$ and $v_t=k(t)^{-1}\sum_{i\in\mathcal{N}_t(k(t))}\bigl(\widetilde S_{t,i}^{(-i)}\bigr)^2$. Choose $\lambda_t$ by \eqref{eq:betting-function}.
    \State Compute $\widetilde{S}_t$ and $S_t$
\[
    \begin{aligned}
\widetilde{S}_t
&=
\frac{
\sum_{i\in\mathcal N_t(k(t))}
\left[
\mathrm{K}(X_i,X_t)
+\mathrm{K}(Y_i,Y_t)
-\mathrm{K}(X_i,Y_t)
-\mathrm{K}(Y_i,X_t)
\right]
}{
\sqrt{2}
\max\left\{
\left(
\sum_{i,j\in\mathcal N_t(k(t))}
\left[
\mathrm{K}(X_i,X_j)
+\mathrm{K}(Y_i,Y_j)
-\mathrm{K}(X_i,Y_j)
-\mathrm{K}(Y_i,X_j)
\right]
\right)^{1/2},
\eta k(t)
\right\}
},\\
S_t&=w(Z_t)\widetilde{S}_t;
\end{aligned}
\]
    \State Update $M_t$:
    \[
    M_{t}=M_{t-1}\times (1+\lambda_tS_t);
    \]
    \If{$M_t > 1/\alpha$}
        \State  reject $H_0$.
    \EndIf
\EndFor
\end{algorithmic}
\end{algorithm}

\subsection{Theoretical Results}\label{theory}
In this section, we study the theoretical properties of SCPT, including anytime-valid Type I error control under the null hypothesis and asymptotic power under the alternative. Proofs of all theoretical results are deferred to Section~S1.1 of the Supplementary Material.

We first establish the anytime-valid Type I error control of SCPT. The result follows directly from the martingale property and Ville's inequality. Notably, its validity does not require the independence condition in Assumption~\ref{assumption_independence}. Therefore, SCPT is anytime-valid for any infinite data stream \(\{(X_s,Y_s,Z_s)\}_{s=1}^{\infty}\) satisfying the null hypothesis in \eqref{H0:orig}. The result allows us to test the null hypothesis continuously as data arrive while maintaining valid overall Type I error control.

\begin{theorem}\label{H0}
Let \(M_t\) denote the process generated by Algorithm~\ref{alg:approx}. Suppose that Assumptions~\ref{assumption_kernel}--\ref{weight} hold. Then, under the null hypothesis in \eqref{H0:orig}, for any \(\alpha\in(0,1)\),
\begin{equation}
\mathbb{P}\left(
\exists\, t\geq 1 \text{ such that } M_t> \alpha^{-1}
\right)
\leq \alpha.
\end{equation}
\end{theorem}

To characterize the asymptotic behavior of $M_t$ under the alternative hypothesis, we impose the following additional conditions, which we briefly discuss in Remark~\ref{comment}.

\begin{assumption}\label{assumption_distribution}
\begin{enumerate}[label=(\roman*)]
\item $\{Z_t\}_{t=1}^{\infty}$ shares a common support $\mathcal{S}$, on which $Z_t$ has a continuous Lebesgue density function $f_t$ for each $t\geq 1$. 
\item  Define the strong mixing coefficient of the covariate sequence $\{Z_t\}_{t=1}^{\infty}$ by
\[
\alpha_Z(m)
=
\sup_{t\geq 1}
\sup_{\substack{
A\in \sigma(Z_1,\ldots,Z_t),\\
B\in \sigma(Z_{t+m},Z_{t+m+1},\ldots)
}}
\left|
\mathbb{P}(A\cap B)-\mathbb{P}(A)\mathbb{P}(B)
\right|.
\]
%We assume that, 
There exists some constant $r\in(0,1)$ such that
\[
\sum_{m=1}^{\infty}\alpha_Z(m)^r <\infty .
\]
\item $K_0= \mathcal{S}\cap\bar{\mathcal{Z}_0}$ is compact, where $\bar{\mathcal{Z}_0}$ denotes the closure of $\mathcal{Z}_0$. Moreover, there exists a constant $c>0$ such that $f_t\geq c$ on $K_0$ for each $t\geq 1$.

\item There exists a random triple $(X,Y,Z)$, with $Z$ supported on $\mathcal{S}$, such that, as $t\to\infty$,
\begin{equation}
\sup_{z\in K_0}
d_{\mathrm{TV}}\left(
\mathbb{P}_{(X_t,Y_t)\mid Z_t=z},
\mathbb{P}_{(X,Y)\mid Z=z}
\right)
\to 0,
\end{equation}
where $d_{\mathrm{TV}}$ denotes the total variation distance between probability measures.
\item For each $t\geq 1$, there exists a measurable subset $\mathcal{Z}_t\subset \mathcal{Z}_0$ such that
\[
\mathbb{P}(Z_t\in \mathcal{Z}_t) \geq c_0,
\quad
\mathrm{MMD}(\mathbb{P}_{X\mid Z=z}, \mathbb{P}_{Y\mid Z=z})\geq c_1
\quad \text{on }  \{z\in\mathcal{Z}_t\},
\]
where $c_0>0, c_1 >0 $ are constants independent of $t$.

\item There exist constants \(c_2>0\) and \(r_0>0\) such that $\left|B(x,r)\cap K_0\right|\geq c_2 \left|B(x,r)\right|$ for all $x\in K_0$ and $0<r<r_0$, where $|\cdot|$ denotes the Lebesgue measure, and $B(x,r)$ denotes the closed Euclidean ball centered at $x$ with radius $r$. 
\item There exists a constant \(C_0>0\) and a continuous function
\(h:\mathcal{S}\times\mathcal{S}\to\mathbb{R}_{+}\) such that, for any \(t\geq 1\) and
\(z,z'\in\mathcal{S}\),
\begin{equation}\label{Lipschitz}
\begin{split}
    d_{\mathrm{TV}}\left(\mathbb{P}_{(X_t,Y_t)\mid Z_t=z}, \mathbb{P}_{(X_t,Y_t)\mid Z_t=z'}\right)
    \leq C_0\left\{1+h(z,z')\right\}\|z-z'\|_2.
\end{split}
\end{equation}

\end{enumerate}
\end{assumption}

\begin{remark}\label{comment} 
Condition (i) is a standard density assumption. Condition (ii) requires the covariate sequence to be strongly mixing, with mixing coefficients satisfying a summability condition. This condition is strictly weaker than independence across time, as it permits temporal dependence that decays sufficiently rapidly as the time gap increases. Our theory therefore applies to a broader class of data streams, allowing covariates with temporal dependence and varying distributions.
Condition (iii) requires $K_0=\mathcal{S}\cap\overline{\mathcal{Z}_0}$ to be compact, which holds automatically when $\mathcal{S}$ is bounded. Even when the support is unbounded, one can choose $\mathcal{Z}_0$ to be a sufficiently large ball or cube covering the covariate values of interest, so that the restriction has little effect on the intended application. Such restrictions are common in conditional nonparametric testing \citep{crump2008nonparametric,chang2015nonparametric,zheng2025conditional}. Nevertheless, compactness serves primarily a technical role in establishing almost sure convergence, and our simulations suggest that SCPT can continue to perform well even when this assumption is not satisfied. 
Conditions (iv) and (v) describe the limiting behavior under the alternative. Broadly, condition (iv) requires the joint conditional distributions to stabilize asymptotically, while condition (v) requires the probability of violating the null hypothesis in \eqref{H0:iid} to be bounded away from zero.  In the stationary setting, one may take $(X,Y,Z)=(X_1,Y_1,Z_1)$ when verifying condition (iv). These conditions simplify the asymptotic analysis and can potentially be relaxed. 
The remaining conditions impose standard regularity requirements. The geometric condition on $K_0$ is similar to those adopted in previous work \citep{jiang2019non,zhao2019minimax}. The local Lipschitz-type condition controls how the conditional distributions vary with the covariates. Related smoothness assumptions on conditional expectations are common in theoretical analyses of KNN methods \citep{azadkia2021simple,huang2022evaluating}. Although regularity of certain conditional expectations would suffice, we impose the condition on the full conditional distribution for simplicity.
\end{remark}

The next theorem characterizes the behavior of $M_t$ under the alternative hypothesis.

\begin{theorem}\label{H1}
Suppose that 
$k(t)\asymp t^\beta$ for some $0<\beta<1$, and Assumptions~\ref{assumption_independence}--\ref{assumption_distribution} hold.
Let $(X,Y,Z)$ denote the random triple in
Assumption~\ref{assumption_distribution}, and define
\[
\Delta(z)
=
\operatorname{MMD}^2
\left(
\mathbb P_{X\mid Z=z},
\mathbb P_{Y\mid Z=z}
\right).
\]
Then we have the following results.
\begin{enumerate}[label=(\roman*)]
\item For any prescribed significance level \(0<\alpha<1\), the
null hypothesis is rejected in finite time almost surely; that is,
\[
\mathbb{P}(\tau_\alpha<\infty)=1,
\]
where
\[
\tau_\alpha
=
\inf\{t\geq 1:M_t> \alpha^{-1}\}.
\]
\item The asymptotic logarithmic growth rate satisfies
\begin{equation}\label{bound_MT_as}
\liminf_{t\to\infty}
\left\{
\frac{\log M_t}{t}
-
\frac{1-\log 2}{2t}
\sum_{s=1}^{t}
\mathbb E\left[
\frac{
w(Z_s)\Delta^2(Z_s)
}{
\max\{\sqrt{\Delta(Z_s)},\eta\}^2
}
\right]
\right\}
\geq0.
\end{equation}
Consequently,
\begin{equation}\label{bound_MT_as_2}
 \liminf_{t\to\infty}
\frac{\log M_t}{t}   \geq \frac{(1-\log 2)w_0 c_0 c_1^4}{2\max\{c_1,\eta\}^2}.
\end{equation}
\end{enumerate}
\end{theorem}

The first result shows that the null hypothesis is rejected in finite time almost surely, and hence the proposed SCPT has asymptotic power one. The second result implies that $M_t$ diverges to infinity almost surely at an exponential rate.

\subsection{Extensions}\label{extension}

We now present several extensions of the standard SCPT procedure presented in Algorithm~\ref{alg:approx}. These address more general dependence structures, as discussed at the beginning of Section~\ref{construction}, and sensitivity to hyperparameter choices. We also consider variants tailored to one-versus-many and CGOF testing.

\subsubsection{More General Dependence Structures}\label{general_dependence}

Under suitable assumptions, SCPT can accommodate temporal
dependence beyond the conditional independence across time
assumed above. In particular, the current responses may depend
on a finite history of covariates and responses. For fixed
nonnegative integers $l_0$ and $l_1$, define the augmented covariate
\[
\widetilde Z_t
=
\left(
Z_t,Z_{t-1},\ldots,Z_{t-l_0},
X_{t-1},\ldots,X_{t-l_1},
Y_{t-1},\ldots,Y_{t-l_1}
\right).
\]
Suppose that the conditional distribution of $(X_t,Y_t)$ given
$(Z_t,\mathcal F_{t-1})$ depends on the past only through the
components included in $\widetilde Z_t$. This finite-order
Markov-type assumption can be written as
\[
\mathbb P_{(X_t,Y_t)\mid Z_t,\mathcal F_{t-1}}
=
\mathbb P_{(X_t,Y_t)\mid\widetilde Z_t}.
\]

To simplify the exposition, consider the global null, with no restriction to a subset of covariate values. Let $\widetilde{\mathcal S}_t$ denote the support of $\widetilde Z_t$. Under the condition above, the null hypothesis in \eqref{H0:1} becomes
\[
H_0:\quad
\mathbb P_{X_t\mid \widetilde Z_t=\widetilde z}
=
\mathbb P_{Y_t\mid \widetilde Z_t=\widetilde z}
\quad
\text{for any } t\geq 1
\text{ and } \widetilde z\in\widetilde{\mathcal S}_t.
\]
The nearest-neighbor constructions in SCPT can then be applied to $\widetilde Z_t$ in place of $Z_t$. In particular, the neighborhood at time $t$ consists of past observations with similar augmented histories:
\[
\mathcal N_t(k(t))
=
\left\{
1\leq s<t:
\widetilde Z_s \text{ is among the } k(t)
\text{ nearest neighbors of } \widetilde Z_t
\right\}.
\]
With this substitution, the conditional mean-zero argument continues to hold under the null, so the resulting process $\{M_t\}_{t\geq 0}$ remains a valid test martingale. Establishing asymptotic power guarantees under this broader dependence structure is more challenging and is beyond the scope of this paper. Nevertheless, the simulations in Section~\ref{simulation} suggest that SCPT performs well in such settings.

\subsubsection{Combining Multiple Hyperparameter Configurations}\label{mixture} 	%Using a Mixture of Betting Factors}\label{mixture}

SCPT involves several prespecified choices, including the kernel,
the neighborhood selection rule, and the weight function $w$.
A practical way to reduce sensitivity to these choices is to
combine the betting factors obtained from multiple hyperparameter
configurations. To illustrate this approach, consider $m$
different configurations. For example, one may vary $a>0$ and $0<b<1$ in the neighborhood-size rule $k(t)=\lceil at^b\rceil$. Let
\[
\lambda_t^{(1)}(Z_t),\ldots,\lambda_t^{(m)}(Z_t)
\quad\text{and}\quad
S_t^{(1)},\ldots,S_t^{(m)}
\]
denote the betting fractions and payoffs constructed under these configurations. We set $M_0=1$ and define the following process by mixing betting factors corresponding to different hyperparameter configurations:
\[
M_t
=
\prod_{s=1}^{t}
\left(
1+\sum_{\ell=1}^{m}
w_{\ell,s}\lambda_s^{(\ell)}(Z_s)S_s^{(\ell)}
\right),
\]
where each $w_{\ell,s}$ is $\mathcal F_{s-1}$-measurable, non-negative, and satisfies
\(
\sum_{\ell=1}^{m}w_{\ell,s}=1.
\)
Under the null hypothesis, predictability of the weights gives
\[
\begin{aligned}
\mathbb E\left[
\sum_{\ell=1}^{m}
w_{\ell,t}\lambda_t^{(\ell)}(Z_t)S_t^{(\ell)}
\;\middle|\; \mathcal F_{t-1}
\right]=
\sum_{\ell=1}^{m}w_{\ell,t}
\mathbb E\left[
\lambda_t^{(\ell)}(Z_t)S_t^{(\ell)}
\;\middle|\; \mathcal F_{t-1}
\right]
=0.
\end{aligned}
\]
 Thus, $\{M_t\}_{t\geq 0}$ remains a valid test martingale and retains the same anytime-valid Type I error control. The simplest choice is $w_{\ell,s}=1/m$, which averages the betting factors across configurations at each time. By concavity of the logarithm, the equal-weight mixture inherits the average of the componentwise lower bounds on asymptotic logarithmic growth and therefore remains consistent whenever each configuration satisfies the conditions of Theorem~\ref{H1}. Alternatively, the weights may be adapted to past observations to place greater emphasis on more informative configurations, potentially improving power \citep{waudby2025universal}. We leave a detailed investigation of such adaptive weighting schemes to future work.

\subsubsection{One-versus-Many and CGOF Testing}\label{cgof_test}

We consider two additional testing problems within the SCPT framework: {one-versus-many} testing and conditional goodness-of-fit (CGOF) testing.

In the one-versus-many setting, at each time $t$ we observe $Z_t$, $Y_t$, and $B$ auxiliary observations $\{X_t^{(b)}\}_{b=1}^B$ that are conditionally 
i.i.d. given $Z_t$.
Suppose that Assumption~\ref{assumption_independence} holds with \((X_t,Y_t)\) being replaced by \((X_t^{(1)},\ldots,X_t^{(B)},Y_t)\). We aim to test the null hypothesis \eqref{H0:iid}, with $X_t$ replaced by $X_t^{(1)}$. This setting arises, for example, in calibration testing when multiple reference observations are available from the target conditional distribution. To incorporate all $B$ auxiliary observations, we compute an SCPT payoff for each paired stream and average the resulting payoffs. Specifically, for each $b=1,\ldots,B$, we apply \eqref{eq:tilde-S} to the stream $\{(X_t^{(b)},Y_t,Z_t)\}_{t\geq1}$ to obtain
$$
\begin{aligned}
\widetilde{S}^{(b)}_t
=
\frac{
\sum_{i\in\mathcal N_t(k(t))}
\left[
\mathrm{K}(X^{(b)}_i,X_t^{(b)})
+\mathrm{K}(Y_i,Y_t)
-\mathrm{K}(X^{(b)}_i,Y_t)
-\mathrm{K}(Y_i,X^{(b)}_t)
\right]
}{
\sqrt{2}
\max\left\{
\left(
\sum_{i,j\in\mathcal N_t(k(t))}
\left[
\mathrm{K}(X^{(b)}_i,X_j^{(b)})
+\mathrm{K}(Y_i,Y_j)
-\mathrm{K}(X^{(b)}_i,Y_j)
-\mathrm{K}(Y_i,X^{(b)}_j)
\right]
\right)^{1/2},
\eta k(t)
\right\}
}.
\end{aligned}
$$
We then define the average payoff $S_t$ by
\[
S_t=w(Z_t)\widetilde{S}_t, \text{~where~}\widetilde{S}_t=\frac{1}{B}\sum_{b=1}^{B}\widetilde{S}^{(b)}_t.
\]
For each $b=1,\ldots,B$ and $i\in\mathcal{N}_t(k(t))$, compute the
leave-one-out payoff $\widetilde S_{t,i}^{(b,-i)}$ by applying
\eqref{eq:loo-payoff} to the stream $\{(X_s^{(b)},Y_s,Z_s)\}_{s\geq 1}$.
For each neighbor $i\in\mathcal{N}_t(k(t))$, averaging the leave-one-out payoffs across the $B$ streams gives
\[
\widetilde S_{B,t,i}^{(-i)}
=\frac{1}{B}\sum_{b=1}^{B}\widetilde S_{t,i}^{(b,-i)}.
\]
Using these averages, we estimate the first two conditional moments of $\widetilde S_t$ by
\[
u_t=\frac{1}{k(t)}\sum_{i\in\mathcal{N}_t(k(t))}\widetilde S_{B,t,i}^{(-i)},
\qquad
v_t=\frac{1}{k(t)}\sum_{i\in\mathcal{N}_t(k(t))}
\bigl(\widetilde S_{B,t,i}^{(-i)}\bigr)^2.
\]
We substitute $u_t$ and $v_t$ into \eqref{eq:betting-function} to obtain $\lambda_t$
and update the test martingale $M_t$ as in Algorithm~\ref{alg:approx}.

In the CGOF setting, the target conditional distribution is specified directly. Suppose that the observed pairs $\{(Y_t,Z_t)\}_{t\geq1}$ are independent across time, and consider the null hypothesis
\[
\mathbb{P}(Y_t\leq u\mid Z_t=z)=F(u\mid z)
\]
for any $t\geq1$, $u\in\mathbb{R}^q$, and $z\in\mathcal S_t\cap\mathcal Z_0$, where $F$ is a known conditional distribution function. Generating a single auxiliary observation $X_t$ from $F(\cdot\mid Z_t)$ at each time reduces this problem to the standard SCPT. Because $F$ is known, we can instead integrate the payoff over the auxiliary observations, with the observed responses and covariates held fixed, to obtain
$$
\begin{aligned}
\widetilde{S}_t
=
\mathbb{E}_{X\mid Z}\left\{
\frac{
\sum_{i\in\mathcal N_t(k(t))}
\left[
\mathrm{K}(X_i,X_t)
+\mathrm{K}(Y_i,Y_t)
-\mathrm{K}(X_i,Y_t)
-\mathrm{K}(Y_i,X_t)
\right]
}{
\sqrt{2}
\max\left\{
\left(
\sum_{i,j\in\mathcal N_t(k(t))}
\left[
\mathrm{K}(X_i,X_j)
+\mathrm{K}(Y_i,Y_j)
-\mathrm{K}(X_i,Y_j)
-\mathrm{K}(Y_i,X_j)
\right]
\right)^{1/2},
\eta k(t)
\right\}
}
\right\},
\end{aligned}
$$
where the expectation $\mathbb{E}_{X\mid Z}$ is taken over independent
draws $X_i\sim F(\cdot\mid Z_i)$, $i=1,\ldots,t$, with the observed
data $\{(Y_i,Z_i)\}_{i=1}^t$ held fixed. We then set
$S_t=w(Z_t)\widetilde{S}_t$.

For each $i\in\mathcal{N}_t(k(t))$, let $\widetilde S_{t,i}^{(-i)}$
denote the single-stream leave-one-out payoff obtained by applying
\eqref{eq:loo-payoff} to the stream $\{(X_s,Y_s,Z_s)\}_{s\geq 1}$.
Here, each $X_s$ is a hypothetical auxiliary draw from
$F(\cdot\mid Z_s)$, and the draws are independent conditional on the
observed data. Integrating over the auxiliary responses gives the CGOF
leave-one-out payoff
\[
\widetilde S_{\infty,t,i}^{(-i)}
=\mathbb E_{X\mid Z}\!\left[\widetilde S_{t,i}^{(-i)}\right].
\]
We estimate the first two conditional moments of $\widetilde S_t$ by
\[
u_t=\frac{1}{k(t)}\sum_{i\in\mathcal{N}_t(k(t))}
\widetilde S_{\infty,t,i}^{(-i)},
\qquad
v_t=\frac{1}{k(t)}\sum_{i\in\mathcal{N}_t(k(t))}
\bigl(\widetilde S_{\infty,t,i}^{(-i)}\bigr)^2.
\]
Substituting $u_t$ and $v_t$ into \eqref{eq:betting-function} gives
$\lambda_t$, and we update $M_t$ as in Algorithm~\ref{alg:approx}. Although the conditional expectations defining the CGOF payoffs can in principle be evaluated exactly, doing so is often infeasible in practice, so we typically approximate them by Monte Carlo.
Conditional on the observed stream $\{(Y_t,Z_t)\}_{t\geq1}$, we draw
$X_t^{(b)}\sim F(\cdot\mid Z_t)$ independently across $t\geq1$ and
$b=1,\ldots,B$, and average both the current payoffs and each neighbor's
leave-one-out payoffs across the resulting $B$ paired streams
$\{(X_t^{(b)},Y_t,Z_t)\}_{t\geq1}$. This yields exactly the
one-versus-many construction described above. Thus, in practice, the
CGOF version of SCPT is typically implemented through a one-versus-many
procedure. 

The theoretical guarantees for the one-versus-many and CGOF extensions can also be established. Specifically, under suitable conditions, both extensions and their equal-weight mixtures described in Section~\ref{mixture} enjoy theoretical guarantees analogous to those established for the standard SCPT algorithm. Moreover, an additional result shows that increasing the number $B$ of auxiliary streams does not reduce asymptotic logarithmic capital growth, and exact CGOF integration performs at least as well as any finite-$B$ version in this sense. This suggests that increasing the number of auxiliary streams can improve test power. The corresponding theoretical results and proofs are provided in Section~S1.2 of the Supplementary Material.

\section{Simulation Studies}\label{simulation}
In this section, we conduct simulation studies to evaluate SCPT in three settings: independent and identically distributed observations, nonstationary or dependent streams, and one-versus-many or conditional goodness-of-fit testing. For each setting, we run 500 independent repetitions, each with at most 4,000 observations, and set the significance level to \(\alpha=0.05\). We use the Dirac kernel for categorical responses and the Gaussian kernel
for continuous responses. For the latter, we select $\sigma$ using the median heuristic based on the first $t_0=50$ observations and hold it fixed thereafter. Sequential testing begins at observation 51 (see Section~S2.1 of the Supplementary Material for further implementation details).

For SCPT, we set $w(Z_t)\equiv1$ and $k(t)=\lceil a\sqrt{t}\rceil$, with $a\in\{2,4,6\}$. In addition to the three fixed-$a$ SCPT procedures, we consider their equal-weight mixture defined in Section~\ref{mixture}, which we denote by SCPT mixture.

\subsection{Independent and Identically Distributed Observations}\label{simulation-iid}
We compare SCPT with a baseline that applies the fixed-sample ECMMD test of \citet{chatterjee2024kernel} with Bonferroni correction across times. More precisely, we perform their test every 50 or 100 observations, denoting the resulting procedures by ECMMD50 and ECMMD100. At the $j$th test, the fixed-sample $p$-value is compared with $\alpha/\{j(j+1)\}$. The first test uses 100 observations.  For both ECMMD50 and ECMMD100, the fixed-sample test at each test uses the asymptotic null distribution derived by \citet[Theorem~4.3]{chatterjee2024kernel} under the condition $K(t)=o(t^{1/44})$, which allows the number of nearest neighbors $K(t)$ to grow only very slowly with the sample size $t$. Therefore, we consider fixed nearest-neighbor counts $K\in\{20,40,60\}$. We report the pointwise maximum empirical rejection rate over $K\in\{20,40,60\}$ for ECMMD in each setting. 

We consider four continuous-response configurations and two additional categorical-response configurations. Across all settings, we use $\gamma\geq 0$ to control the discrepancy between the two conditional distributions, with $\gamma=0$ corresponding to the null hypothesis and larger values indicating greater discrepancies. For continuous responses, we generate
\[
Z\sim U(-1,1)^{10},\qquad
u(Z)=\frac{1}{\sqrt{10}}\sum_{j=1}^{10}Z_j,\qquad
\varepsilon_X,\varepsilon_Y\stackrel{\mathrm{i.i.d.}}{\sim}N(0,1),
\]
where the two noise variables are independent of $Z$. 

\medskip
\noindent\textit{Homoscedastic Mean Difference.}\,
We generate
\[
X=u(Z)+\varepsilon_X,\qquad
Y=(1+\gamma)u(Z)+\varepsilon_Y.
\]
Both conditional variances are one, and the conditional means differ by $\gamma u(Z)$.

\medskip
\noindent\textit{Homoscedastic Variance Difference.}\,
We generate
\[
X=u(Z)+\varepsilon_X,\qquad
Y=u(Z)+\exp(\gamma)\varepsilon_Y.
\]
The conditional means coincide, whereas the conditional variances are $1$ and $\exp(2\gamma)$.

\medskip
\noindent\textit{Heteroscedastic Mean Difference.}\,
We generate
\[
X=u(Z)+\exp\{u(Z)\}\varepsilon_X,\qquad
Y=(1+\gamma)u(Z)+\exp\{u(Z)\}\varepsilon_Y.
\]
Both conditional variances equal $\exp\{2u(Z)\}$, and the conditional mean difference is $\gamma u(Z)$.

\medskip
\noindent\textit{Variance Difference with One Homoscedastic Group.}\,
We generate
\[
X=u(Z)+\varepsilon_X,\qquad
Y=u(Z)+\exp\{\gamma u(Z)\}\varepsilon_Y.
\]
The conditional means coincide. The conditional variance of $X$ is one, while that of $Y$ is $\exp\{2\gamma u(Z)\}$. 

\medskip
\noindent\textit{Binary Classification.}\,
We generate a predicted success probability $Z\sim\mathrm{Beta}(2,8)$ and, conditionally on $Z$, draw
\[
X\mid Z\sim\mathrm{Bernoulli}(Z)\text{~and~}
Y\mid Z\sim\mathrm{Bernoulli}\{q_\gamma(Z)\},
\]
independently, where
\[
q_\gamma(z)
=z+\frac{\gamma}{2}
\left\{
\text{expit}\left(\frac{0.1-z}{0.02}\right)
-\text{expit}\left(\frac{z-0.9}{0.02}\right)
\right\},\qquad
\text{expit}(v)=\frac{1}{1+\exp(-v)}.
\]
This perturbation increases small success probabilities and decreases large ones, while leaving intermediate probabilities nearly unchanged. Such a pattern can arise when risk-prediction models overfit and produce overly extreme predictions. 

\medskip

\noindent\textit{Three-Class Classification.}\,
We generate a predicted probability vector
\[
Z=(Z_1,Z_2,Z_3)\sim\mathrm{Dir}(4,4,4),
\]
and draw $X$ and $Y$ independently given $Z$, with
\[
\mathbb P(X=j\mid Z)=Z_j,\qquad
\mathbb P(Y=j\mid Z)=
\frac{Z_j^{\,1+\gamma}}
{\sum_{\ell=1}^{3}Z_\ell^{\,1+\gamma}},
\quad j=1,2,3.
\]
Increasing $\gamma$ concentrates the observed class probabilities toward the more likely classes.

\medskip

\begin{figure}[tbp]
    \centering
    \includegraphics[width=0.75\linewidth]{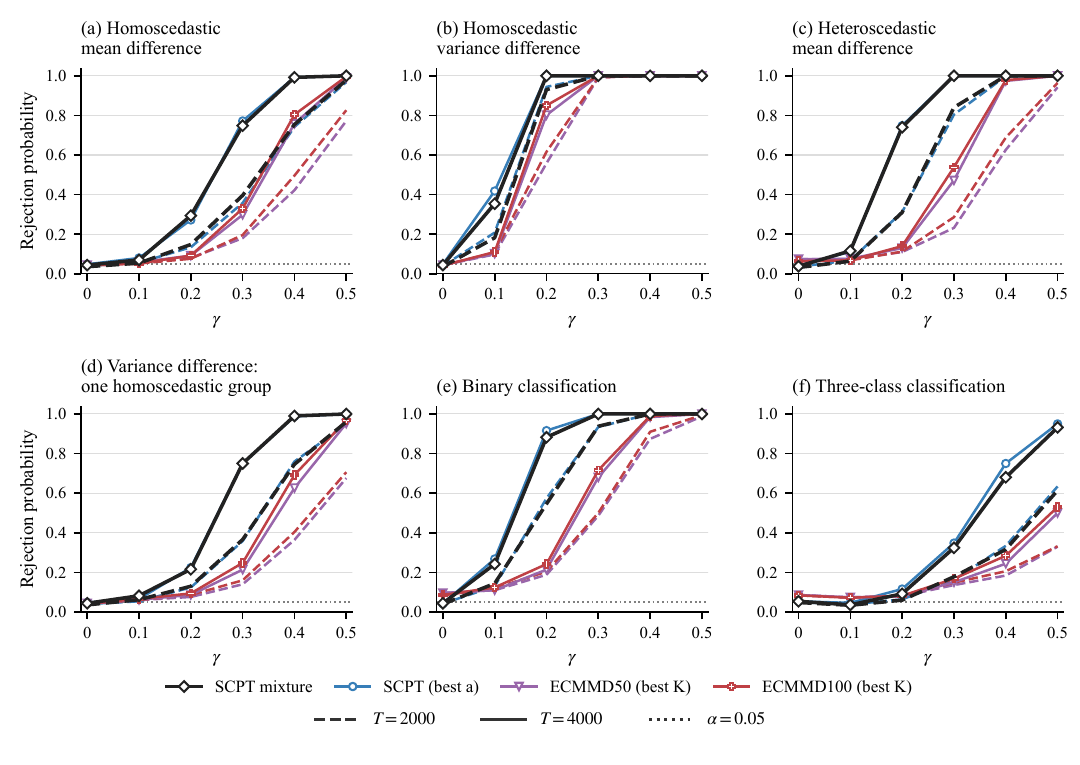}
    \caption{Empirical rejection rates for independent and identically distributed observations. Panels (a)--(d) are the four continuous-response settings; panels (e)--(f) are binary classification and three-class classification. The null occurs at $\gamma=0$. The SCPT mixture uses equal weights over $a\in\{2,4,6\}$, and SCPT (best \(a\)) shows the pointwise maximum empirical rejection rate over \(a\in\{2,4,6\}\). ECMMD50 and ECMMD100 show their respective pointwise maximum empirical rejection rates over $K\in\{20,40,60\}$. Dashed and solid curves show rejection by $T=2000$ and $T=4000$, respectively. The horizontal dotted line marks significance level $\alpha=0.05$. }
    \label{fig:simulation-iid}
\end{figure}

Figure~\ref{fig:simulation-iid} reports empirical rejection rates by times $T=2000$ and $T=4000$. Under the null ($\gamma=0$), the maximum empirical Type I error rates across the six settings by $T=4000$ are $5.4\%$ for the SCPT mixture, $5.4\%$ for the individual SCPT procedures, $9.6\%$ for ECMMD50, and $8.6\%$ for ECMMD100. Under the alternatives ($\gamma>0$), the SCPT mixture generally outperforms the Bonferroni procedures. For example, at $\gamma=0.2$ and $T=4000$, its rejection rate in the heteroscedastic mean setting is $74.0\%$, compared with $14.0\%$ for ECMMD100 and $13.0\%$ for ECMMD50. The corresponding rates in the binary setting are $88.2\%$, $24.2\%$, and $21.4\%$. These gains are obtained even though each baseline is allowed retrospective selection of its best $K$. 

We also note that the SCPT mixture achieves rejection rates close to those obtained by selecting the best $a\in\{2,4,6\}$ retrospectively for each alternative. Among the 18 alternatives with moderate to large discrepancies ($\gamma\geq0.3$), the largest relative shortfall at $T=4000$ is $9.3\%$. It occurs in the three-class setting at $\gamma=0.4$, where the mixture and the best individual procedure have rejection rates of $68.0\%$ and $75.0\%$, respectively. The mixture matches the benchmark in 13 of the 18 settings and slightly exceeds it in the variance-difference setting with one homoscedastic group at $\gamma=0.4$ ($99.0\%$ versus $98.8\%$). Thus, the mixture achieves broadly competitive performance using a single prespecified combination.

\subsection{Nonstationary and Dependent Streams}\label{simulation-dependent}

We next consider four nonstationary settings and two settings with temporal dependence. For nonstationary or dependent streams, tests such as ECMMD, whose critical values rely on asymptotic null distributions derived under the i.i.d.\ assumption, are no longer guaranteed to control Type I error. We therefore focus on SCPT, whose sequential conditional formulation accommodates these settings. For the four nonstationary settings, we retain the independent
ten-dimensional uniform covariates used in the i.i.d. settings.
We consider abrupt change and gradual drift under both homoscedastic
and heteroscedastic noise. Let $\varepsilon_{X,t}$ and
$\varepsilon_{Y,t}$ be independent standard normal variables,
independent across time and of the covariates. In the homoscedastic
settings, we generate
\[
X_t=u(Z_t)+\varepsilon_{X,t},\qquad
Y_t=\{1+\gamma d_t\}u(Z_t)+\varepsilon_{Y,t}.
\]
In the heteroscedastic settings, we generate
\[
X_t=u(Z_t)+\exp\{u(Z_t)\}\varepsilon_{X,t},\qquad
Y_t=\{1+\gamma d_t\}u(Z_t)+\exp\{u(Z_t)\}\varepsilon_{Y,t}.
\]
In all four settings, the observations are independent across time.
The conditional mean difference at $Z_t=z$ is $\gamma d_t u(z)$,
where the deterministic sequence $d_t$ specifies how the discrepancy
evolves over time.

\medskip
\noindent\textit{Abrupt Change.}\,
We set
\[
d_t=\mathbb I\{t>500\}.
\]
That is, an abrupt conditional mean change is introduced after time 500.

\medskip
\noindent\textit{Gradual Drift.}\,
We set
\[
d_t=\min\left\{1,\max\left\{0,\frac{t-500}{1000}\right\}\right\}.
\]
The mean difference increases linearly after time 500, reaches its full magnitude at time 1500, and then remains constant.

\medskip

In the two dependent settings, $X_t$, $Y_t$, and $Z_t$ are scalar-valued. Let
$\{\varepsilon_t^Z\}$, $\{\varepsilon_t^X\}$, and
$\{\varepsilon_t^Y\}$ be mutually independent sequences of
i.i.d.\ $N(0,1)$ variables, and define
\[
Z_t=0.1\sum_{j=1}^{5}Z_{t-j}+\varepsilon_t^Z,
\qquad
w_t=\frac{1}{\sqrt6}\sum_{j=0}^{5}Z_{t-j}.
\]

\medskip
\noindent\textit{Dependence through Covariate Histories.}\,
We generate
\[
X_t=w_t+\varepsilon_t^X,\qquad
Y_t=(1+\gamma)w_t+\varepsilon_t^Y,
\]
and construct nearest neighbors using
\[
\widetilde Z_t=(Z_t,Z_{t-1},\ldots,Z_{t-10})\in\mathbb R^{11}.
\]
The dependence of the current conditional distributions on the past is entirely represented by the covariate history.

\medskip
\noindent\textit{Dependence through Covariate and Response Histories.}\,
We define
\[
m_t=w_t+0.04\sum_{j=1}^{5}(X_{t-j}+Y_{t-j})
\]
and generate
\[
X_t=m_t+\varepsilon_t^X,\qquad
Y_t=m_t+\gamma w_t+\varepsilon_t^Y.
\]
SCPT uses
\[
\widetilde Z_t=
\bigl(
Z_t,\ldots,Z_{t-10},
X_{t-1},\ldots,X_{t-10},
Y_{t-1},\ldots,Y_{t-10}
\bigr)\in\mathbb R^{31}.
\]
The two groups have the same history feedback and conditional variance, and their conditional means differ by $\gamma w_t$. At $\gamma=0$, the conditional distributions of $X_t$ and $Y_t$ coincide given the full past and current covariate.

\medskip

\begin{figure}[htbp]
    \centering
    \includegraphics[width=0.75\linewidth]{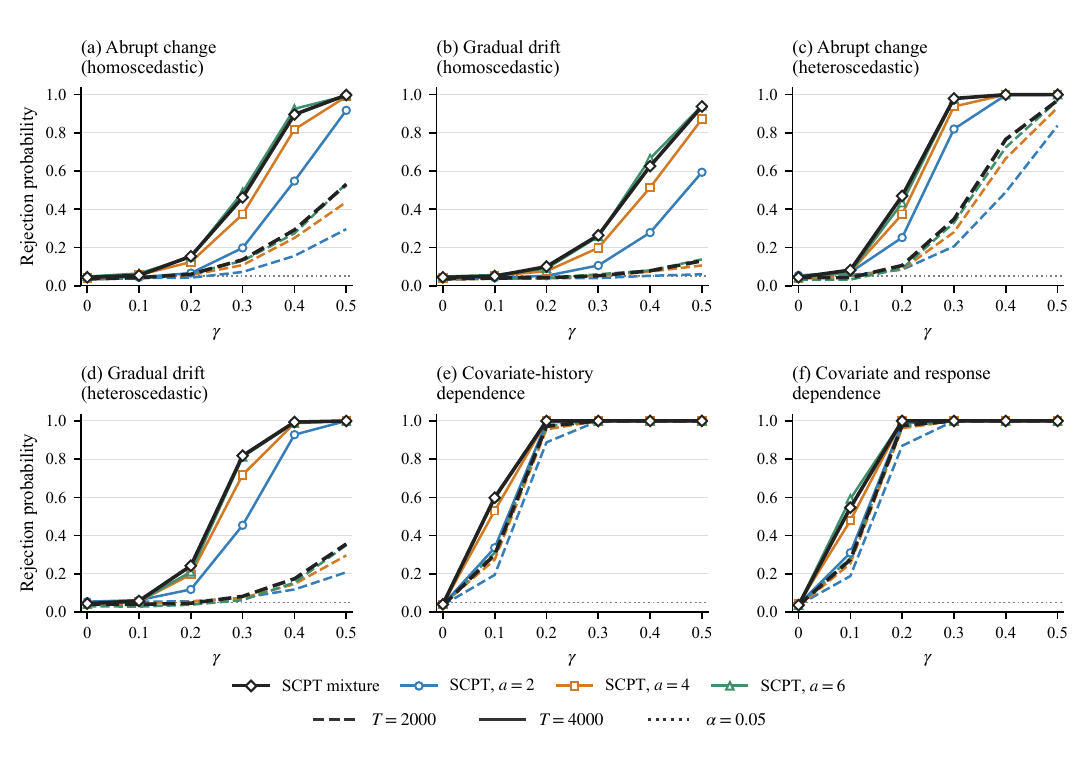}
    \caption{Cumulative rejection rates for nonstationary and dependent streams. Panels (a)--(b) show abrupt change and gradual drift under homoscedastic noise. Panels (c)--(d) show the same changes under heteroscedastic noise. In all four settings, the change begins after time 500. Panels (e)--(f) show dependence through covariate histories and through covariate and response histories, respectively, with true order five and augmentation to order ten. All panels display SCPT with $a=2,4,6$ and their equally weighted mixture. The null occurs at $\gamma=0$. Dashed and solid curves correspond to $T=2000$ and $T=4000$; the dotted horizontal line marks significance level $\alpha=0.05$.}
    \label{fig:simulation-dependent}
\end{figure}

We apply the extension described in
Section~\ref{general_dependence} to the two dependent settings.
The true dependence order is five. To reflect a conservative
choice when this order is unknown, SCPT uses histories
extending to lag ten. Figure~\ref{fig:simulation-dependent} reports cumulative rejection rates by times $T=2000$ and $T=4000$ for the six nonstationary and dependent settings. Under the null ($\gamma=0$), the maximum empirical Type I error rates across these settings by $T=4000$ are $4.4\%$ for the SCPT mixture and $5.4\%$ for the individual SCPT procedures.

For the four nonstationary settings, the cumulative rejection rate includes both false alarms at or before time 500 ($\tau\leq500$) and first rejections after the change ($500<\tau\leq T$). At $\gamma=0.5$ and $T=4000$, the SCPT mixture has cumulative rejection rates of $99.8\%$, $93.8\%$, $100\%$, and $100\%$ in panels (a)--(d), respectively. In each homoscedastic setting, $2.4\%$ of the 500 repetitions give a false alarm by time 500, compared with $1.8\%$ in each heteroscedastic setting. The proportions of all repetitions that first reject after the change and by $T=4000$ are therefore $97.4\%$, $91.4\%$, $98.2\%$, and $98.2\%$, respectively. 

The SCPT mixture remains close to the best individual SCPT procedure across these different patterns. For homoscedastic abrupt change at $\gamma=0.4$ and $T=4000$, it achieves a cumulative rejection rate of $89.6\%$, compared with $92.6\%$ for the best individual procedure. Under heteroscedastic gradual drift at $\gamma=0.3$ and $T=4000$, the corresponding rates are $81.8\%$ and $81.2\%$. The prespecified mixture again achieves rejection rates comparable to those obtained by selecting the best $a$ retrospectively for each alternative.

\subsection{One-versus-Many and CGOF Testing}\label{simulation-extension} We evaluate the one-versus-many and CGOF extensions of SCPT
described in Section~\ref{cgof_test}. As discussed there,
the conditional expectations defining the CGOF payoff often
cannot be evaluated exactly in practice and can instead be
approximated by averaging over finitely many auxiliary streams.
We therefore consider SCPT with 1, 10, and 20 auxiliary streams,
denoted by SCPT1, SCPT10, and SCPT20, respectively, without
separately evaluating the exact CGOF procedure.

We compare these procedures with the kernel conditional Stein
discrepancy test (KCSD), a fixed-sample CGOF test proposed by
\citet{jitkrittum2020testing}. We apply KCSD sequentially using
the same Bonferroni spending rule, and refer to the two resulting procedures as
KCSD50 and KCSD100, respectively.
KCSD requires access to the conditional score function of the
specified density and is supplied with its exact form in these
experiments, whereas SCPT requires only sampling access to
the model. We use 1000 multinomial bootstrap draws to calculate
the KCSD $p$-value. 
The bootstrap justification for KCSD assumes i.i.d.\ observations.
In contrast, SCPT does not require i.i.d.\ observations and
provides finite-sample, anytime-valid Type I error control
under the relevant conditional null. It can also accommodate
covariate drift and temporal dependence under the assumptions
of the framework.

We consider the following two model families adapted from
\citet{jitkrittum2020testing}.

\medskip
\noindent\textit{Heteroscedastic Gaussian Model.}\,
We generate $Z\sim N(0,I_3)$ and specify the target and observation models as
\[
\begin{aligned}
X\mid Z
&\sim N\left(\sum_{j=1}^{3}Z_j,v_\gamma(Z)\right),
\qquad
Y\mid Z
\sim N\left(\sum_{j=1}^{3}Z_j,1\right),\\
v_\gamma(Z)
&=1+\gamma\exp\left\{
-\frac{\|Z-c\|_2^2}{2\times(0.8)^2}
\right\},
\qquad
c=\frac23(1,1,1)^{\top}.
\end{aligned}
\]
The target variance differs from the observation variance
primarily near $c$, producing a localized discrepancy.

\medskip
\noindent\textit{Quadratic Gaussian Model.}\,
We generate $Z\sim U(-1,1)$ and set
\[
X\mid Z\sim N(Z+1,1),\qquad
Y\mid Z\sim N(Z+1+\gamma Z^2,1).
\]
The target model is linear, whereas the observation model has
an additional quadratic term. This mimics a practical scenario
where the model is too simplistic to fit the data adequately.

\medskip

\begin{figure}[htbp]
    \centering
    \includegraphics[width=0.7\textwidth]{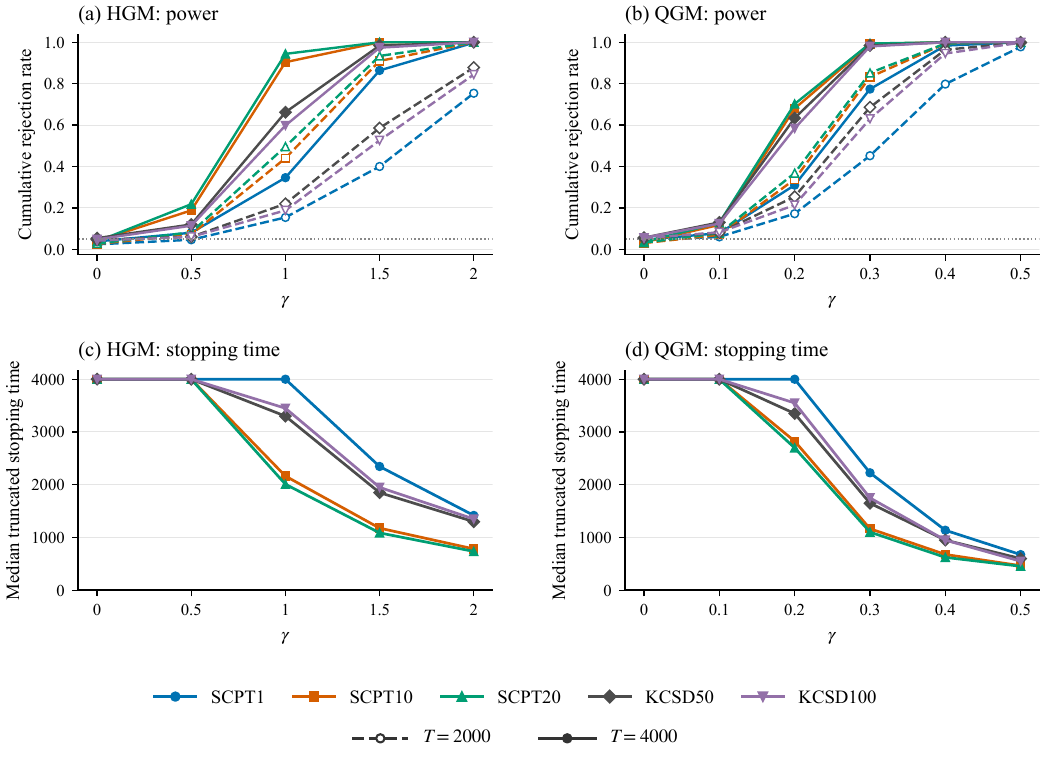}
    \caption{Comparison of SCPT with Bonferroni-adjusted KCSD. The left and right columns
    show the heteroscedastic and quadratic Gaussian models,
    respectively.
    The upper panels show cumulative rejection probabilities
    at 2000 observations (dashed lines with open markers)
    and 4000 observations (solid lines with filled markers).
    The lower panels show median truncated stopping times.
    SCPT1, SCPT10, and SCPT20 use 1, 10, and 20 auxiliary
    streams, respectively, and combine the betting factors
    for $a\in\{2,4,6\}$ with equal weights.
    KCSD50 and KCSD100 use 1000 bootstrap draws and the
    Bonferroni spending rule.
    Dotted horizontal lines in the upper panels mark
    the nominal level $\alpha=0.05$.}
    \label{fig:simulation-cgof}
\end{figure}

Figure~\ref{fig:simulation-cgof} compares cumulative rejection
probabilities at 2000 and 4000 observations and the median
truncated stopping time,
\(
\operatorname{median}\{\min(\tau,4000)\},
\)
assigning 4000 to repetitions without rejection. Under the null ($\gamma=0$), the maximum empirical Type I error rates across the two model families by $T=4000$ are $4.6\%$ for SCPT1, $4.0\%$ for SCPT10, $3.8\%$ for SCPT20, and $5.6\%$ for both KCSD50 and KCSD100. In the heteroscedastic Gaussian setting, SCPT10 and SCPT20
show clear advantages in both power and rejection speed at
moderate discrepancy levels. For example, when $\gamma=1$,
SCPT10 achieves rejection probabilities of $44.0\%$ and
$90.4\%$ at 2000 and 4000 observations, respectively, compared
with $22.0\%$ and $66.2\%$ for KCSD50 and $18.8\%$ and
$59.8\%$ for KCSD100. Its median truncated stopping time is
2164, compared with 3300 and 3450 for KCSD50 and KCSD100. In the quadratic Gaussian setting, the power advantage of
SCPT10 and SCPT20 is smaller,
but their median truncated stopping times show a
pronounced improvement. For example, when $\gamma=0.3$,
the rejection probabilities at 4000 observations are
$99.2\%$ for SCPT10, $98.4\%$ for KCSD50, and $98.0\%$
for KCSD100, while their median truncated stopping times
are $1166$, $1650$, and $1750$, respectively.
Thus, even when the methods have similar power at the final
horizon, the one-versus-many SCPT typically requires fewer observations to
reject the null hypothesis. Across both model families, increasing the number of auxiliary
streams from one to ten noticeably improves performance.
Increasing it further to twenty provides little additional gain, suggesting that a moderate number of auxiliary streams already capture
much of the benefit of averaging.

Overall, the simulations illustrate the power and flexibility of SCPT in paired conditional two-sample testing, including nonstationary and dependent settings, as well as one-versus-many and CGOF testing. The mixture extension in Section~\ref{mixture} provides a simple way to combine betting factors across different neighborhood rules, while averaging payoffs over auxiliary streams can improve power when multiple auxiliary observations
are available at each time point. The Supplementary Material (Sections~S3.1--S3.5) provides more detailed summaries of median truncated stopping times and cumulative rejection rates for all simulation settings, along with additional results for higher-dimensional ($p=100$) covariates. These experiments provide further evidence of SCPT's advantages in the settings considered.

\section{Real-Data Applications}\label{real-data}
In this section, we apply SCPT to two online testing problems using real data. In the first, we compare readings from co-located environmental sensors. In the second, we assess the goodness of fit of conditional density models. Throughout, we set the significance level to $\alpha=0.05$ and use an equally weighted factor mixture over nearest-neighbor coefficients $a\in\{2,4,6\}$, as in the simulation study.

\subsection{Comparing Temperature and Humidity Measurements}
\label{sec:real-sensors}

We compare measurements for temperature and relative humidity from a low-cost AHT10 sensor and a professional CS215-L reference sensor from the microclimate dataset provided by \citet{schoeneberger2026multisensor}. The data were collected near Gr\"unstadt, Germany, in June 2025. The two instruments were installed in the same radiation shield at a height of 2\,m. The original study compares readings from the two sensors and reports \(R^2=0.9992\) and an RMSE of \(0.18\,\mathrm{K}\) for temperature, and \(R^2=0.9720\) and an RMSE of \(2.64\) percentage points for relative humidity.

We analyze temperature and relative humidity measurements separately, taking $X_t$ and $Y_t$ to be the corresponding readings from the AHT10 and CS215-L sensors. $Z_t$ consists of incoming shortwave and longwave radiation, both measured in $\mathrm{W/m^2}$ by a separate SN-500 radiometer. These variables describe the radiative conditions shared by the two instruments. We apply SCPT to monitor whether the two instruments’ measurement distributions agree given the shared environmental covariate \(Z_t\) and the observed past.

\begin{figure}[tbp]
\centering
\includegraphics[width=\textwidth]{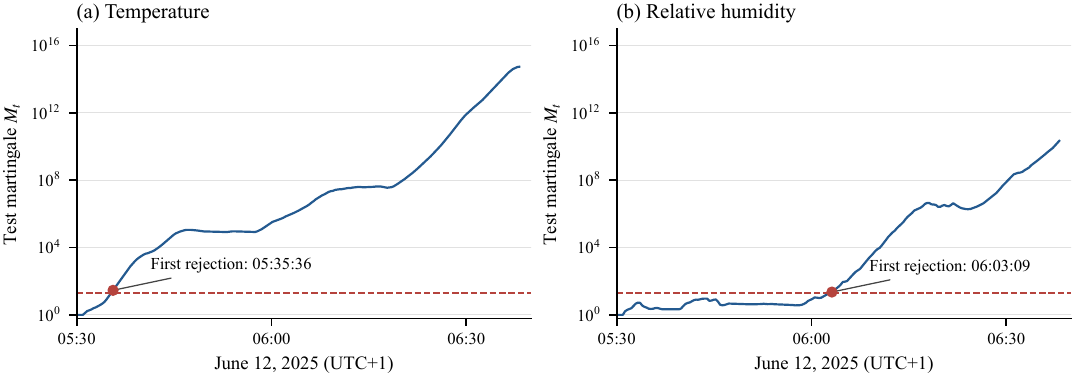}
\caption{SCPT test martingale curves for co-located temperature and relative-humidity measurements, with logarithmic vertical axes. Panels (a) and (b) show temperature and relative humidity, respectively, from 05:30 to 06:40 on June 12, 2025. Dashed horizontal lines indicate the rejection threshold $1/\alpha=20$ at significance level $\alpha=0.05$, and marked points indicate first rejection.}
\label{fig:real-sensor-capital}
\end{figure}

We sequentially test the null hypothesis defined in~\eqref{H0:1} using data from June 12--14 in chronological order. We select the Gaussian-kernel bandwidth by the median heuristic using the first 50 pairs, as in the simulation studies. Monitoring begins after the first 50 pairs. Figure~\ref{fig:real-sensor-capital} shows the SCPT test martingales during the initial monitoring period.  On June~12, the temperature test first rejects at the 15th monitoring observation (05:35:36), and the humidity test first rejects at the 90th (06:03:09). Both martingales show further growth after rejection. Although readings from the two instruments are highly correlated, SCPT can still promptly detect differences in their conditional distributions given the shared environmental conditions and the observed past.

\subsection{New York City Taxi}\label{real-data-taxi}
We consider online monitoring of conditional density models for taxi drop-off locations using New York City yellow-taxi records from January 2015. Following existing studies \citep{dutordoir2018gaussian,jitkrittum2020testing}, we restrict our analysis to trips with both pick-up
and drop-off locations in or near Manhattan. Section~S2.3 of the Supplementary Material provides further details of the analysis.

We compare two Gaussian mixture density networks fitted using data from January 1--20. The first model conditions only on the pick-up location; the second model additionally incorporates time of day and day of week. SCPT10 uses ten destinations sampled from the fitted conditional model to test the model's goodness-of-fit. For both experiments, we order the trips recorded from January 21 onward by pick-up time and retain every 100th trip. We use the 3,823 trips retained on January 21 to select the kernel bandwidth via the median heuristic. Starting on January 22, we apply the sequential test to the retained trips.

\begin{figure}[tbp]
\centering
\includegraphics[width=1.0\textwidth]{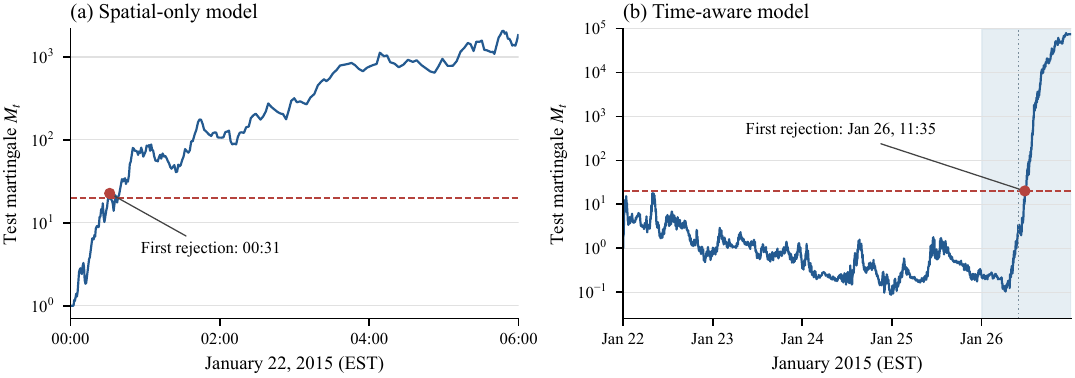}
\caption{SCPT10 test martingale curves for the two taxi models, with logarithmic vertical axes. Panel (a) shows the spatial-only model from midnight to 06:00 on January 22; panel (b) shows the time-aware model from January 22 through the end of January 26. Dashed horizontal lines indicate the rejection threshold $1/\alpha=20$ at significance level $\alpha=0.05$, and marked points indicate first rejection. Shading in panel (b) marks January 26; the dotted vertical line marks the 09:51 observation of moderate snow at Central Park. Trip times refer to pick-up.}
\label{fig:real-taxi-capital}
\end{figure}

\medskip
\noindent\textit{A Model without Temporal Covariates.}\,
For the spatial-only model, SCPT10 rejects at 00:31 on January 22, after 53 monitoring observations. Figure~\ref{fig:real-taxi-capital}(a) shows the SCPT10 test martingale through 06:00. The martingale continues to grow after rejection. The rejection is consistent with temporal variation in trip destinations that a model conditioning only on pick-up location cannot capture. Table~\ref{tab:taxi-morning} compares observed and model-predicted mean
straight-line trip distances on January 22. By the time SCPT10 first
rejects at 00:31:32, the observed mean over the first 53 trips is 3.076\,km, compared with the model's prediction of
2.407\,km. This early-morning gap persists over the full
00:00--06:00 interval, with observed and predicted means of 3.021
and 2.467\,km, respectively. During 06:00--18:00, the pattern
reverses: the observed mean falls to 2.213\,km, below the predicted
mean of 2.392\,km. The observed and predicted mean distances are close over the full day (2.420\,km and 2.397\,km, respectively), but the model underestimates trip distances in the early morning and overestimates them during the daytime.

\begin{table}[ht]
\centering
\small
\caption{Observed and model-predicted mean straight-line trip distances on January 22. Distances are measured in kilometers. All times refer to trip pick-up times}
\label{tab:taxi-morning}
\begin{tabular}{lrrr}
\toprule
Time interval & Trips & Observed & Predicted \\
\midrule
00:00--00:31:32 (first rejection) & 53 & 3.076 & 2.407 \\
00:00--06:00 & 247 & 3.021 & 2.467 \\
06:00--18:00 & 2,246 & 2.213 & 2.392 \\
18:00--24:00 & 1,499 & 2.630 & 2.392 \\
\midrule
Full day & 3,992 & 2.420 & 2.397 \\
\bottomrule
\end{tabular}
\end{table}

\medskip
\noindent\textit{A Model with Temporal Covariates:}\,
With time of day and day of week included, SCPT10 remains below the rejection threshold from January 22 through January 25. Its test martingale rises sharply on the morning of January 26 and first crosses the threshold at 11:35, after 16,956 monitoring observations. Figure~\ref{fig:real-taxi-capital}(b) shows the process through the end of January 26. The absence of the earlier alarm is consistent with improved modeling of regular temporal variation. The later growth of the test martingale on January 26 coincides with the progression of the January 26--27 New York winter storm, from moderate snowfall in the morning to heavier snowfall in the afternoon and evening. In particular, the Central Park station recorded moderate snow and visibility of half a mile at 09:51 EST on January 26 \citep{iem2015centralpark}. This temporal alignment is consistent with snow-related changes in the conditional destination distribution.

\section{Discussion}\label{conclusion}

In this work, we developed SCPT, a sequential nonparametric procedure for testing the equality of conditional distributions in paired data streams with shared covariates. SCPT combines kernel mean embeddings, nearest-neighbor estimation, and the test-by-betting framework. Under the null, the conditional mean-zero payoff yields a non-negative martingale and hence finite-sample, anytime-valid Type I error control. Under suitable regularity conditions, we establish consistency through a strictly positive lower bound on the asymptotic logarithmic growth rate. This analysis accommodates certain time-varying distributions and strongly mixing covariates, extending beyond the i.i.d.\ setting commonly considered in the theoretical analysis of anytime-valid inference \citep{shekhar2023nonparametric,chugg2023auditing,podkopaev2023sequential,arnold2026betting,cho2026gaavi}. SCPT can also be extended to conditional two-sample testing
with separate covariate observations, although constructing
a valid test martingale in this setting requires additional
assumptions. When the ratio of the two covariate densities is known,
this extension can be achieved by reweighting the observations.
We describe the construction in Section~S5 of the Supplementary
Material, where we also discuss its connection to the model-X
framework for sequential conditional independence testing
\citep{grunwald2024anytime}.

SCPT is also related to existing two-sample tests. When the observations are independent across time, the global null
in \eqref{H0:orig} reduces to equality of the conditional distributions
of $X_t$ and $Y_t$ given $Z_t$. Since the two responses
share $Z_t$, this is equivalent to equality of the joint distributions
of $(X_t,Z_t)$ and $(Y_t,Z_t)$. This suggests applying the sequential two-sample test of
\citet{shekhar2023nonparametric} to the joint observations. However, a generic joint two-sample test may fail
to exploit the conditional structure and consequently suffer
a loss of power when discrepancies are localized
in covariate space, as discussed by \citet{jin2026testing}
in the i.i.d.\ setting. Our supplementary experiments
(Section~S4 of the Supplementary Material) show that the joint
two-sample betting procedure has substantially lower power
than SCPT in such settings.

The nearest-neighbor construction, similar to those used by \citet{azadkia2021simple} and \citet{chatterjee2024kernel}, allows SCPT to estimate conditional kernel mean embeddings without specifying a model for the conditional distributions. It nevertheless inherits the curse of dimensionality associated with nonparametric regression, which can slow the accumulation of evidence and lead to longer stopping times when the covariates are high-dimensional. For fixed-sample testing, \citet{jin2026testing} proposed an alternative based on generative models and cross-fitting to handle high-dimensional covariates. Extending this approach to sequential testing, however, is not straightforward. Other potential approaches include dimensionality reduction,
learned representations, and data-adaptive distance metrics
for nearest-neighbor selection. Incorporating these approaches
may improve performance with high-dimensional covariates and
warrants further investigation.

A further limitation concerns the dependence structures covered
by our asymptotic theory. Although our analysis under the alternative assumes
conditional independence across time given the covariates,
numerical results suggest that SCPT can perform well under
broader forms of dependence, including the finite-order Markov
setting described in Section~\ref{general_dependence}. However, establishing
consistency in such settings remains challenging.
Extending the theory to more general dependence
structures, together with developing methods for high-dimensional
covariates, would broaden the applicability of SCPT to complex
streaming-data problems.

\section*{Acknowledgements} 

This work was supported by the National Natural Science Foundation of China (Grant No. 12671339) and the Anhui Center for Applied Mathematics.

\bibliographystyle{apalike}
\bibliography{ref}

\end{document}